\documentclass[a4paper,11pt]{article}
\pdfoutput=1

\usepackage{jcappub}
\usepackage[utf8]{inputenc} 
\usepackage{amsfonts,amssymb,mathrsfs,amsmath,esint,bm}
\usepackage{pdflscape} 
\usepackage[capitalise]{cleveref}
\usepackage{hyperref}
\allowdisplaybreaks

\graphicspath{{./Figures/}}
\usepackage{latexsym}
\usepackage{graphicx}
\usepackage[dvipsnames]{xcolor}
\usepackage{booktabs}
\usepackage{physics}
\usepackage{tikz}
\usetikzlibrary{decorations.pathmorphing}\usetikzlibrary{positioning,calc}
\usetikzlibrary{math}
\usepackage{cleveref}
\usepackage{comment}
\usepackage[normalem]{ulem}
\newcommand{\stkout}[1]{\ifmmode\text{\sout{\ensuremath{#1}}}\else\sout{#1}\fi} 
\graphicspath{{./Figures/}}

\renewcommand{\d}{\mathrm{d}}

\renewcommand{\epsilon}{\varepsilon}

\newcommand*\diff{\mathop{}\!\mathrm{d}}

\def\bk{\boldsymbol{k}}

\def\bx{\boldsymbol{x}}

\definecolor{poleCross}{rgb}{0,0.8,.13}
\definecolor{poleRH}{rgb}{0,0.84,1}

\title{
Nonminimal Superheavy Dark Matter
}
\author[a]{Sarunas~Verner}

\affiliation[a]{Institute for Fundamental Theory, Physics Department, University of Florida,\\ Gainesville, FL 32611, USA}

\emailAdd{verner.s@ufl.edu}

\abstract{
We investigate the gravitational production of superheavy scalar fields with nonminimal coupling during and after inflation. We derive analytical approximations using the mode function solution in a de Sitter background and also apply the steepest descent method. We study both positive and negative nonminimal couplings and show that their comoving number density spectra behave differently in the short and long wavelength regimes. We numerically compute the comoving number density spectra and dark matter abundance for a broad range of superheavy spectator fields exceeding the Hubble scale during inflation, with $m_{\chi} > H_I$, and nonminimal couplings ranging from $-300 \leq \xi \leq 300$. When computing the allowed dark matter parameter space, we impose the maximum reheating temperature constraint, the Big Bang nucleosynthesis constraint, and the isocurvature constraint. We show that the presence of a positive or negative coupling $\xi$ can expand the parameter space up to $3$ orders of magnitude above the Hubble inflationary scale, allowing such dark matter candidates to be as heavy as $\sim 10^{16} \, \rm{GeV}$ in the Starobinsky model of inflation.}

\begin{document}

\maketitle

\section{Introduction}
\label{sec:introduction}
The profound nature and origin of dark matter (DM) continue to pose significant challenges to our understanding of fundamental physics, despite its existence being inferred from Fritz Zwicky's observations of the Coma Cluster $90$ years ago~\cite{Zwicky:1933gu}. Modern cosmology, rich with precision data from cosmic microwave background (CMB) measurements and large-scale structure (LSS) surveys, confirms that the universe contains significantly more dark matter than visible matter. However, the stringent constraints imposed by current direct detection searches, such as XENON1T~\cite{XENON:2018voc}, LUX~\cite{LUX:2016ggv}, PandaX~\cite{PandaX-II:2020oim}, and LZ~\cite{LZ:2022lsv}, coupled with the absence of indirect detection, challenge the standard weakly interacting massive particle (WIMP) paradigm and motivate the exploration of minimalistic alternative models~\cite{Arcadi:2017kky, Escudero:2016gzx, Arcadi:2024ukq}.

The observed inhomogeneities on cosmological scales have led to the development of theories such as cosmic inflation, characterized by the rapid, accelerated expansion of the early universe. Some models postulate that the dark sector was populated independently from the visible sector during inflation and the post-inflationary reheating epoch,\footnote{For a review on inflation, see Refs.~\cite{Lyth:1998xn, Senatore:2016aui, Baumann:2018muz}} through mechanisms such as the freeze-in process~\cite{Hall:2009bx, Bernal:2017kxu}. This process proposes that dark matter could have been produced early in the universe and subsequently decoupled from the thermal bath, thereby evading current direct detection bounds while matching the current DM relic abundance $\Omega_{\rm DM} h^2 = 0.1198 \pm 0.0012$, as measured by the \textit{Planck} observations of the CMB~\cite{Planck:2018vyg}. This scenario is further supported by the possibility of non-perturbative gravitational particle production, which suggests that dark sector particles (spectator fields) were produced gravitationally during the transition from a quasi-de Sitter inflationary phase to a matter- or radiation-dominated epoch~\cite{Parker:1968mv, Parker:1969au, Ford:1986sy, Ema3, Chung:1998rq}. Typically, these models impose stringent constraints on the isocurvarture power spectrum from the CMB, implying that the produced spectator scalar field must be superheavy, with a mass comparable to the Hubble inflationary scale~\cite{Chung:2004nh, Chung:2011xd, Chung:2015pga, Herring:2019hbe, Padilla:2019fju, Ling:2021zlj, Redi:2022zkt}.

Recently, gravitational particle production has garnered significant attention, leading to extensive studies in various contexts~(see Ref.~\cite{Kolb:2023ydq} for a review). This process has been thoroughly investigated for dark matter models with different spins: spin-0~\cite{Chung:1998zb, Chung:1998ua, Kolb:1998ki, Ling:2021zlj, Garcia:2023qab, Racco:2024aac}, spin-$\frac{1}{2}$~\cite{Lyth:1996yj, Kuzmin:1998kk, Chung:2011ck}, spin-1~\cite{Graham:2015rva, Ahmed:2020fhc, Kolb:2020fwh,Gorghetto:2022sue, Cembranos:2023qph,Ozsoy:2023gnl}, spin-$\tfrac 32$ \cite{Hasegawa:2017hgd,Antoniadis:2021jtg, Kaneta:2023uwi,Casagrande:2023fjk}, and spin-2 \cite{Kolb:2023dzp}. Furthermore, gravitational production during inflation inevitably contributes to the dark sector, as outlined in~\cite{Gross:2020zam, Redi:2020ffc, Krnjaic:2020znf, Arvanitaki:2021qlj, Redi:2022zkt, Redi:2022myr, East:2022rsi, Bastero-Gil:2023htv, Bastero-Gil:2023mxm}. This production occurs both during and after inflation. The onset of the universe's thermal history during the reheating epoch can further lead to graviton-mediated particle creation via inflaton condensate or thermal Standard Model (SM) bath scattering~\cite{Greene:1998nh, Ema:2015dka, Garny:2015sjg, Ema:2016hlw, Tang:2017hvq, Garny:2017kha, Bernal:2018qlk, Ema:2018ucl, Bettoni:2018utf, Garny:2018grs, Opferkuch:2019zbd, Bettoni:2019dcw, Chianese:2020yjo, Herring:2020cah, Mambrini:2021zpp, Bernal:2021kaj, Barman:2021ugy, Garani:2021zrr, Bettoni:2021zhq, Ahmed:2021fvt, Haque:2021mab, Clery:2021bwz, Haque:2022kez, Aoki:2022dzd, Clery:2022wib, Garcia:2022vwm, Kaneta:2022gug, Ahmed:2022tfm, Basso:2022tpd, Barman:2022qgt, Lebedev:2022vwf, Haque:2023yra, Zhang:2023xcd, Laverda:2023uqv, Zhang:2023hjk, Figueroa:2024asq, Racco:2024aac, Choi:2024bdn}.

A comprehensive analytical and numerical study of minimally-coupled superheavy dark matter was presented in~\cite{Racco:2024aac}, employing the steepest descent method~\cite{Chung:1998bt, Enomoto:2020xlf} for realistic inflation models with a time-varying Hubble parameter. In typical plateau-like inflationary models, such as the Starobinsky model~\cite{Starobinsky:1980te} or the T-model of inflation~\cite{Kallosh:2013hoa}, the occupation number for long-wavelength infrared (IR) modes is exponentially suppressed when $m_{\chi} \gg H_I$, with $n_k \propto \exp(-\alpha m_{\chi}/H_I)$, where $\alpha \lesssim 2\pi$. This aligns with predictions from a Bose-Einstein distribution with the Gibbons-Hawking temperature $T_{\rm dS} = H_I/2\pi \ll m_{\chi}$~\cite{Gibbons:1977mu}, implying that $\omega_k/T_{\rm dS} \sim 2\pi m_{\chi}/H_I$ and that gravitational scalar dark matter production in de Sitter space is given by $n_k \simeq e^{-2\pi m_{\chi}/H_I}$\cite{Markkanen:2017rvi}. For short-wavelength ultraviolet (UV) modes, the occupation number has an exponentially suppressed $k$-dependent tail, $n_k \propto \exp(-c k^2)$, where $c$ is a model-dependent constant.

In this work, we extend the study to consider the production of nonminimally-coupled superheavy dark matter during and after inflation.\footnote{One may even be tempted to call such dark matter candidates nonminimal WIMPzillas.} We examine a superheavy spectator scalar field $\chi$ coupled to gravity nonminimally through the interaction term $\xi \chi^2 R$, where $\xi$ represents a dimensionless nonminimal coupling constant that may be positive or negative, and $R$ denotes the Ricci scalar~\cite{Cosme:2017cxk, Cosme:2018nly, Alonso-Alvarez:2018tus, Fairbairn:2018bsw, Alonso-Alvarez:2019ixv, Kolb:2022eyn, Garcia:2023qab}. This term emerges naturally due to the renormalizability of curved spacetime, making $\xi$ a running parameter that is non-zero at all energy scales~\cite{Parker:2009uva, Birrell:1982ix}. As explored in~\cite{Garcia:2023qab}, for light scalars with masses $m_{\chi} < H_I$, nonminimal coupling can either enhance or reduce the gravitationally produced particle abundance. It was shown that for light scalars with a large nonminimal coupling $\xi \gg 1$, gravitational particle production is significantly enhanced, and stringent isocurvature constraints are avoided, allowing for very light dark matter with masses as low as $10^{-4}$ eV, well below the Hubble inflationary scale. This can be understood intuitively by considering the time-dependent mode frequency $\omega_k^2 = k^2 + a^2m_{\chi}^2 + a^2(1/6-\xi)R$. During inflation, the Ricci scalar is approximately $R \simeq -12H_I^2$, and for a large nonminimal coupling, low-$k$ modes are strongly suppressed because the effective mass becomes substantial, with $m_{\rm eff}^2 \simeq 12\xi H_I^2$, thus effectively avoiding bounds on the amplitude of the isocurvature power spectrum~\cite{Chung:2004nh, Chung:2011xd, Chung:2013sla, Chung:2015pga, Markkanen:2017rvi, Herring:2019hbe, Garcia:2023awt}. When the transition from a quasi-de Sitter to a matter- or radiation-dominated universe occurs, it causes the effective mass to vary rapidly, leading to significant particle production. After inflation, during reheating, the oscillating Ricci scalar $R$ leads to strong parametric resonance, which increases with larger values of $\xi$. Consequently, particle production for light scalars with large nonminimal coupling is dominated by short-wavelength (UV) modes in the particle spectra.

\begin{figure}[t!]
\centering
\includegraphics[width=0.98\textwidth]{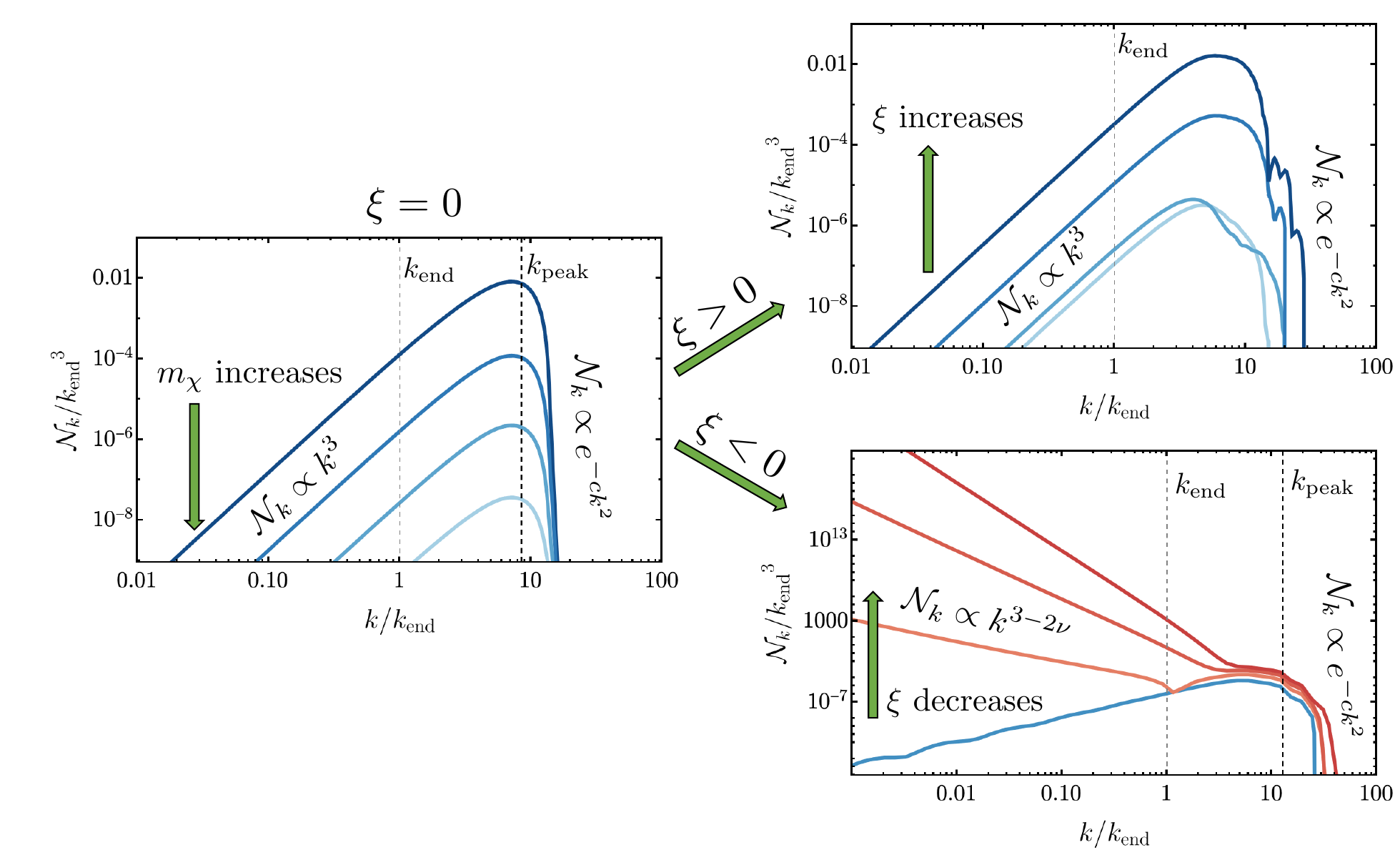}
\caption{A qualitative depiction of the comoving number density spectrum $\mathcal{N}_k$ as a function of $k/k_{\rm end}$, where $k_{\rm end} = a_{\rm end} H_{\rm end}$ represents the mode reentering the horizon at the end of inflation. In the left panel, the behavior of $\mathcal{N}_k$ for $\xi = 0$ is illustrated. Here, as the bare mass $m_{\chi}$ increases, the particle spectrum decreases, with the IR modes scaling as $\mathcal{N}_k \propto k^3$ and the UV modes being exponentially suppressed. The top right panel displays the spectrum for $\xi > 0$, showing an increase as $\xi$ grows, with IR scaling as $\mathcal{N}_k \propto k^3$ and an exponentially suppressed, yet broader and more populated UV tail. The bottom right panel shows the case for $\xi < 0$, where the spectrum increasingly red-tilts as $\xi$ decreases, with IR scaling as $\mathcal{N}_k \propto k^{3-2\nu}$, becoming more dominated by IR modes, although the UV tail remains exponentially suppressed.}
\label{fig:general}
\end{figure}

As a result, at the end of inflation during the reheating phase, parametric resonance driven by oscillations becomes extremely efficient, similar to scenarios with lighter masses. However, the key difference lies in the bare mass of superheavy spectator fields, which results in less efficient parametric resonance. This reduced efficiency can be compensated by a larger nonminimal coupling. It can be shown analytically and numerically that the comoving number density spectrum in the long-wavelength (IR) regime scales as $\mathcal{N}_k \propto k^{3 - 2\nu}$, where $\nu = \sqrt{9/4 - 12\xi - m_{\chi}^2/H_I^2}$ for real $\nu$. For superheavy dark matter with positive $\xi$, when $\nu$ becomes imaginary, the IR spectrum scales as $\mathcal{N}_k \propto k^3$, dominated by short-wavelength UV modes. As $\xi$ increases, the UV peak grows, and the spectrum in the UV becomes noisier due to interference, with exponential suppression at large values of $k$.

In contrast, when $\xi$ is negative, significant negative values lead to a negative mode frequency $\omega_k$, triggering tachyonic resonance during inflation, which results in substantial IR divergence and particle production. Because the spectrum in the IR scales as $\mathcal{N}_k \propto k^{3 - 2\nu}$, the spectrum becomes red-tilted and IR-dominated. Additionally, in this case, the IR divergence leads to a large isocurvature power spectrum and backreaction effects.
With increasing $\xi$, the spectrum at high-$k$ broadens and eventually decreases exponentially for large values of $k$. This behavior, including exponential suppression of the UV tail, mirrors that observed when $\xi \geq 0$. The qualitative behavior of the particle spectra for $\xi = 0$, $\xi > 0$, and $\xi < 0$ is summarized in Fig.~\ref{fig:general}.

The structure of this paper is as follows. In Section~\ref{sec:gravprod},  we briefly review gravitational particle production and discuss the dynamics during inflation. Then, in Section~\ref{sec:generalspectra}, we analyze the particle spectra for nonminimally-coupled superheavy scalar fields, utilizing the steepest descent method for analytical approximations. We explore the production regimes for $\xi = 0$, $\xi > 0$, and $\xi < 0$ in Section~\ref{sec:production}. We calculate the dark matter abundance for nonminimal superheavy dark matter in Section~\ref{sec:dmabundance}, and our conclusions are presented in Section~\ref{sec:conclusions}. Appendix~\ref{app:cosmictime} provides the equations of motion in cosmic time.

\textit{Notation and Conventions}. In this work, we use the natural units $c = \hbar = k_B= 1$ and denote the reduced Planck mass by $M_P \equiv 1/\sqrt{8 \pi G_N} \simeq 2.435 \times 10^{18} \, \rm{GeV}$, where $G_N$ is Newton's constant. 
We use the metric signature $(+, -, -, -)$ and the flat Minkowski spacetime metric corresponds to $\eta_{\mu \nu} dx^{\mu} dx^{\nu} = dt^2 - d {\bf{x}}^2$.

\section{Gravitational Particle Production}
\label{sec:gravprod}
The phenomenon of gravitational particle production emerges due to the expansion of the universe. During the periods of inflation and reheating, fields coupled to gravity experience nonadiabatic mass variations driven by the evolving scale factor (curvature). This results in the excitation of the field's momentum modes due to the evolving background. At late times, these excited modes can be interpreted as a nonzero particle number density. Importantly, the presence of nonminimal coupling, $\xi$, modifies the field's interaction with gravity, leading to various effects during inflation and reheating, which we explore in detail in this paper.

In this section, we introduce our model and outline the procedure for calculating the number and energy density of particles produced by gravitational effects. We then analyze the dynamics of inflation and specify the parameters used in our numerical analysis.
\subsection{Model}
\label{sec:model}
We consider a homogeneous and isotropic background characterized by the Friedmann-Robertson-Walker (FRW) spacetime metric:
\begin{equation}
    \label{eq:frwmetric}
    ds^2 \; = \; a(\eta)^2 \left(d \eta^2 - d \mathbf{x}^2 \right) \, ,
\end{equation}
where $a(\eta)$ is the scale factor, $\mathbf{x}$ denotes a comoving spatial $3$-dimensional vector, and $d \eta \equiv dt/a$ is the conformal (comoving) time. The action of our theory is given by
\begin{equation}
    \label{eq:action}
    \mathcal{S} \; = \; \int d^4 x \sqrt{-g} \left[-\frac{1}{2}\left(M_P^2 - \xi \chi^2 \right)R + \frac{1}{2} \partial_{\mu} \phi \partial^{\mu} \phi - V(\phi) + \frac{1}{2} \partial_{\mu} \chi \partial^{\mu} \chi - \frac{1}{2} m_{\chi}^2 \chi^2 \right] \, ,
\end{equation}
where $g = \mathrm{det}~g_{\mu \nu}$ is the metric determinant, $\xi$ denotes the nonminimal coupling, $\phi$ is the inflaton field, $V(\phi)$ is the inflationary potential, $\chi$ is a spectator scalar field, and $m_{\chi}$ is the bare mass of the spectator scalar. The $Z_2$ symmetry ensures the stability of the spectator field, $\chi$. The choice of $\xi = 0$ (minimal coupling) corresponds to Einstein gravity, while the conformal coupling is given by $\xi = 1/6$. In this work, we primarily use conformal time, $\eta$, with the equations of motion in cosmic time provided in Appendix~\ref{app:cosmictime}.

From this action, the equation of motion for the spectator field can be expressed as
\begin{equation}
    \label{eq:eom1}
    \left[ \frac{d^2}{d\eta^2} - \nabla^2 - \frac{a''}{a} + a^2 m_{\chi}^2 - a^2 \xi R \right] X \; = \; \left(\Box + a^2 m_{\chi}^2 + \frac{a^2}{6} \left(1 - 6 \xi \right) R \right) X \; = \; 0 \, ,
\end{equation}
where we introduced the rescaled field $X \equiv a \chi$. Here, the prime denotes a derivative with respect to conformal time, and in the second line, we express the Ricci scalar as $R  = -6a''/a^3 = -12 H^2 - 6H'/a$, where $H(\eta) = a'/a^2$ is the Hubble parameter, and introduce the d'Alembertian operator $\Box \equiv \partial_{\eta}^2 - \nabla^2$. Since the FRW metric~(\ref{eq:frwmetric}) is homogeneous, one can use the following Fourier decomposition of the spectator field, $X$:
\begin{equation}
\label{eq:modedef}
X(\eta,\bx)  = \int \frac{\diff^3\bk}{(2\pi)^{3/2}}\,e^{-i\bk\cdot\bx} \left[ X_k(\eta)\hat{a}_{\bk} + X_k^*(\eta)\hat{a}^{\dagger}_{-\bk} \right]\,, 
\end{equation}
where $\bk$ is the comoving momentum, with $|\bk| = k$, $\hat{a}_{\bk}$ is the annihilation operator, and $\hat{a}^{\dagger}_{\bk}$ is the creation operator. These operators obey the usual canonical commutation relations $[\hat{a}_{\bk},\hat{a}^{\dagger}_{\bk'} ] = (2\pi)^3 \delta^3(  \bk-\bk')$ and $[\hat{a}_{\bk},\hat{a}_{\bk'} ] = [\hat{a}^{\dagger}_{\bk},\hat{a}^{\dagger}_{\bk'} ] = 0$. The mode functions satisfy the condition $X_k(\eta) = X_{-k}(\eta)$, ensuring the reality condition $X(\eta, \mathbf{x}) = X(\eta, \mathbf{x})^*$. One can show that the canonical commutation relations between the field $X_k$ and its conjugate momentum $X_k'$ implies the Wronskian condition:
\begin{equation}
    \label{eq:wronskian}
    X_k X^{*\prime}_k - X_k^*X_k' \; = \; i \, .
\end{equation}

Substituting the Fourier-decomposed field~(\ref{eq:modedef}) into the equation of motion~(\ref{eq:eom1}), we obtain the equation of motion:
\begin{equation}
    \label{eq:modes}
    X_k''(\eta) + \omega_k^2(\eta) X_k(\eta) \; = \; 0 \, ,
\end{equation}
where the mode frequency is given by
\begin{equation}
    \label{eq:modefreq}
    \omega_k^2(\eta) \; = \; k^2 + a^2(\eta) m^2_{\rm{eff}}(\eta), \quad {\rm{where}} \quad m^2_{\rm{eff}}(\eta)  \; = \; m_{\chi}^2 + \left(\frac{1}{6} -  \xi \right) R(\eta) \, .
\end{equation}
We note that the Ricci scalar term $R$ varies during background evolution, leading to a changing effective mass, which in turn results in the gravitational production of the spectator scalar field.
\subsection{Particle Occupation Number and Energy Density}
\label{sec:numberdensity}
To solve the mode equations~(\ref{eq:modes}), we use the positive frequency of the Bunch-Davies vacuum for the initial condition
\begin{equation}
    \label{eq:bdvacuum}
    X_k(\eta_i) \; = \; \frac{1}{\sqrt{2 \omega_k}} \, , \qquad X_k'(\eta_i) \; = \; - \frac{i \omega_k}{\sqrt{2 \omega_k}} \, ,
\end{equation}
where $|\eta_i \omega_k| \gg 1$ and $\eta_i$ denotes the initial time. Here, the normalization factor of $1/\sqrt{2 \omega_k}$ is necessary to ensure that the canonical commutation relations are satisfied. In the early-time asymptotic limit, $\eta_i \rightarrow -\infty$, this condition implies a flat Minkowski background, where $a^2 R \rightarrow 0$ and $a \rightarrow 0 $, resulting in the mode frequency~(\ref{eq:modefreq}) approaching the limit $\omega_k \rightarrow k$. 

The comoving particle number density is given by\footnote{The Bogoliubov coefficient  $n_k = |\beta_k|^2$ is equivalent to the phase space distribution (PSD) $f_{\chi}(k, \eta)$ used in the Boltzmann equation for number density. See~\cite{Nurmi:2015ema, Garcia:2020wiy, Garcia:2022vwm, Kaneta:2022gug} for a detailed discussion.}
\begin{equation}
    \label{eq:comovingden}
    n_{\chi}(\eta) a(\eta)^3 \; = \; \frac{1}{(2\pi)^3} \int_{k_0}^{\infty} d^3 \bk \,  n_k \; = \; \int_{k_0}^{\infty} \frac{dk}{k} \mathcal{N}_k, \quad {\rm{with}} \quad \mathcal{N}_k \; = \; \frac{k^3}{2\pi^2} |\beta_k|^2 \, ,
\end{equation}
where
\begin{equation}
    n_k \; = \; |\beta_k|^2 \; = \; \frac{1}{2\omega_k} |\omega_k X_k - i X_k'|^2 \, .
\end{equation}
Here, $\mathcal{N}_k$ represents the comoving number density spectrum, and $\beta_k$ is the Bogoliubov coefficient, which is zero at $\eta_i$ and at late times corresponds to the particle occupation number $n_k(\eta) = |\beta_k|^2$. In this expression, we have excluded the vacuum energy contribution of $-\frac{1}{2}$ to ensure that the comoving number density is renormalized and does not contain an ultraviolet (UV) divergence at sufficiently late times.\footnote{This is achieved through the adiabatic regularization scheme or renormalization with normal ordering where rotated Bogoliubov operators are used to eliminate the vacuum energy contribution. See Refs.~\cite{Parker:1974qw, Fulling:1974zr, Ford:1986sy, Anderson:1987yt, Ford:2021syk, Kolb:2023ydq} for a study on regularization.} We also introduced the present-day comoving momentum, $k_0 = a_0 H_0$, assuming that this scale was inside the horizon at the start of inflation and use this quantity as an infrared (IR) cutoff of our integrals, where $a_0$ and $H_0$ are the present-day scale factor and Hubble parameter, respectively~\cite{Starobinsky:1994bd, Ling:2021zlj, Herring:2019hbe}.

Next, we introduce the stress-energy tensor of the spectator scalar field, $\sqrt{-g}T_{\mu \nu,\chi} = 2 \delta(\sqrt{-g} \mathcal{L}_{\chi})/\delta g^{\mu \nu}$~\cite{Birrell:1982ix}, which leads to
\begin{equation}
\begin{aligned}
    T_{\mu \nu}^{\chi} & = \left(1 - 2 \xi \right) \left(\nabla_{\mu} \chi \right) \left(\nabla_{\nu} \chi \right) + \frac{1}{2} \left(4\xi - 1 \right) g_{\mu \nu}g^{\rho \sigma} \left(\nabla_{\rho} \chi \right) \left(\nabla_{\sigma} \chi \right) \\ 
    &- 2 \xi \left(\chi \nabla_{\mu} \partial_{\nu}\chi - g_{\mu \nu} \chi \Box{\chi} \right) + \xi \left[R_{\mu \nu } - \frac{1}{2} R g_{\mu \nu} \right] \chi^2 + \frac{1}{2}g_{\mu \nu} m_{\chi}^2  \chi^2 \, ,
    \label{eq:tmunuscalar}
\end{aligned}
\end{equation}
where $\Box \equiv g_{\mu \nu} \nabla^\mu \nabla^\nu$ and $\nabla^\mu$ represents the covariant derivative in curved spacetime, $R_{\mu \nu}$ is the Ricci tensor, and here the derivatives are taken with respect to cosmic time. The energy density $\rho_{\chi} \equiv T_{00}^{\chi}$ is given by
\begin{equation}
\begin{aligned}
    \rho_{\chi}(\eta) a(\eta)^4 & =  \int \frac{d^3 k }{(2 \pi)^3}  \frac{1}{2}\bigg[ |X_k'|^2 + (k^2 + a^2 m_{\chi}^2)|X_k|^2 +   \\
   &(1 - 6\xi) \left(a^2 H^2 - \frac{1}{6} a^2 R  \right)|X_k|^2 - (1-6\xi) a H (X_k X_k'^* + X_k^* X_k' ) \bigg] \, .
\end{aligned}
\end{equation}
This result simplifies in the late-time limit as $a H \rightarrow 0$ and $a^2 R \rightarrow 0$, and we find that the renormalized energy density of the spectator field, $\chi$, is\footnote{For a discussion on UV divergence regularization, see~\cite{Birrell:1982ix, Kolb:2023ydq}
}
\begin{equation}
    \label{eq:endenrenorm}
    \rho_{\chi}(\eta) a(\eta)^4 \; = \; \frac{1}{(2\pi)^3} \int_{k_0}^{\infty} d^3 \bk \, \omega_k n_k \, .
\end{equation}
We apply these expressions when numerically evaluating the number and energy densities of the spectator scalar field.
\subsection{Inflation}
\label{sec:inflation}
Next, we discuss the inflationary dynamics. From action~(\ref{eq:action}), the inflaton equation of motion can be expressed as
\begin{equation}
    \phi'' + 2 a H \phi' + a^2 \frac{dV(\phi)}{d\phi} \; = \; 0 \, ,
\end{equation}
where $V(\phi)$ is the inflationary potential. Assuming that the background dynamics are solely determined by the motion of the inflaton field, $\phi$, from the Friedmann equation we obtain
\begin{equation}
    \rho_{\phi} \; \equiv \; \frac{\phi'^2}{2a^2} + V(\phi)  \; = \; 3 H^2 M_P^2  \, ,
\end{equation}
where $\rho_{\phi}$ denotes the inflaton energy density. The end of inflation, defined at time $a_{\rm end} \equiv a(\eta_{\rm end})$, occurs when $d(1/aH)/d\eta = 0$ and the comoving Hubble scale ceases to decrease and begins to increase immediately after the end of inflation. Alternatively, the end of inflation is defined when $V(\phi_{\rm end}) = \phi'^2_{\rm end}/a_{\rm end}^2$ and $\rho_{\rm end} \equiv \rho(a_{\rm end}) = \frac{3}{2}V(\phi_{\rm end})$. The Hubble parameter at the end of inflation is denoted as $H_{\rm end} \equiv H(a_{\rm end})$. In the slow-roll approximation, the number of $e$-folds between the Hubble crossing of CMB modes and the end of inflation is expressed as
\begin{equation}
    \label{eq:efolds}
    N_* \; \simeq \; \frac{1}{M_P^2} \int_{\phi_{\rm end}}^{\phi_*} \frac{V(\phi)}{V'(\phi)} d\phi \, ,
\end{equation}
where $k_* = 0.05 \, \rm{Mpc}^{-1}$ is the CMB pivot scale used in the analysis of \textit{Planck}.

When performing the numerical analysis, we use the Starobinsky model of inflation~\cite{Starobinsky:1980te, Ellis:2013nxa, Kallosh:2013yoa},
\begin{equation}
    \label{eq:modestarobinsky}
    V(\phi) \; = \; \frac{3}{4} \lambda M_P^4 \left(1 - e^{-\sqrt{\frac{2}{3}} \frac{\phi}{M_P} } \right)^2 \, ,
\end{equation} 
where $\lambda$ is the normalization scale determined from CMB measurements~\cite{Planck:2018vyg, Planck:2018jri}. The Hubble parameter during inflation can be determined from $H_I \simeq \frac{1}{2}\sqrt{\lambda}M_P$. After the end of inflation, the inflationary potential near the minimum is expressed as as
\begin{equation}
    \label{eq:infpotmin}
    V(\phi) \; \simeq \; \frac12 m_\phi^2 \phi^2 \, , \qquad \phi \lesssim M_P \, ,
\end{equation}
with $m_\phi = \sqrt{\lambda} M_P = 2H_I$. In the large $N_*$ limit, the amplitude of the curvature power spectrum, $A_{S}$, scalar tilt, $n_s$, and tensor-to-scalar ratio, $r$, for the Starobinsky model of inflation are given by
\begin{equation}
    A_{S*} \; \simeq \; \frac{\lambda N_*^2}{24 \pi^2} \, , \qquad n_s \; \simeq \; 1 - \frac{2}{N_*} \, , \qquad r \; \simeq \; \frac{12}{N_*^2} \, ,
\end{equation}
where $A_{S*} \simeq 2.1 \times 10^{-9}$~\cite{Planck:2018jri}.

For a choice of $55$ $e$-folds,\footnote{For simplicity of the analysis, we assume $55$ $e$-folds. For a detailed analysis of the Starobinsky model that constrains the number of $e$-folds while incorporating the effects of reheating, see~\cite{Ellis:2021kad}.} the Starobinsky model of inflation gives $\lambda \simeq 1.6 \times 10^{-10}$, $m_{\phi} \simeq 3 \times 10^{13} \, \rm{GeV}$, $\phi_{\rm end} \simeq 0.61M_P$, $\phi_* \simeq 5.35 M_P$, $H_{\rm end} \simeq 7.5 \times 10^{12} \, \rm{GeV}$, and $H_I \simeq 1.5 \times 10^{13} \, \rm{GeV}$. The scalar tilt is given by $n_s \simeq 0.965$ and the tensor-to-scalar ratio is $r \simeq 0.0035$, in perfect agreement with current \textit{Planck} constraints. We note that for numerical runs, we use the Starobinsky model of inflation with the given parameters. However, the general characteristics of our results, discussed in the following sections, are general and can be applied to various models of inflation.
\section{Particle Spectra}
\label{sec:generalspectra}
In this section, we explore the general properties of the power spectra $\mathcal{P}_{\chi}$ and the comoving number density spectrum $\mathcal{N}_k$ for superheavy dark matter, with $m_{\chi} \gtrsim H_I$. We also investigate the effects of nonminimal coupling and use the steepest descent method to derive analytical approximations for the particle occupation number, $n_k = |\beta_k|^2$.
\subsection{de Sitter Solution and Power Spectrum}
\label{sec:desitter}
To understand the general behavior of gravitationally-produced particle spectra, we first consider the de Sitter solution. In a de Sitter universe, characterized by a constant Hubble parameter $H$, the scale factor and the Ricci scalar are expressed as $a = -\frac{1}{\eta H}$ and $R = -12 H^2$, respectively. By solving the mode equation~(\ref{eq:modes}), we obtain the general solution
\begin{equation}
    \label{eq:gensolxk}
    X_k(\eta) \; = \; c_1 \sqrt{-\eta} H_{\nu}^{(1)} (-k \eta) + c_2 \sqrt{-\eta} H_{\nu}^{(2)}(-k \eta) \, ,
\end{equation}
where $c_1$ and $c_2$ are arbitrary constants, and $H_{\nu}^{(1)}$ and $H_{\nu}^{(2)}$ represent the Hankel functions of the first and second kind, respectively, and
\begin{equation}
    \label{eq:nuterm}
    \nu^2 \; \equiv \; \frac{9}{4} - 12 \xi - \frac{m_{\chi}^2}{H^2} \, .
\end{equation}
Imposing the Wronskian condition~(\ref{eq:wronskian}) and setting $c_2 \rightarrow 0$ from the Bunch-Davies initial condition, the general solution~(\ref{eq:gensolxk}) simplifies to
\begin{equation}
    \chi_k(\eta) \; = \; \frac{\sqrt{\pi}}{2\sqrt{H}} \left(-\eta H \right)^{3/2} e^{i\left(\frac{\pi \nu}{2} + \frac{\pi}{4} \right)} H_{\nu}^{(1)}\left(\frac{k}{aH} \right) \, ,
\end{equation}
where we used $\chi_k = X_k/a$. In the long-wavelength (IR) regime, where $k \ll a H$, we find~\cite{Kolb:2023dzp, Garcia:2023awt}:
\begin{equation}
    \label{eq:modenu}
    \chi_k(k \ll a H) \simeq \frac{1}{2 \sqrt{\pi H}} \left(-\eta H \right)^{3/2} e^{i \frac{\pi}{2}\left(\nu-\frac{1}{2}\right)}\left(e^{-i \pi \nu} \Gamma(-\nu)\left(\frac{k}{2 a H}\right)^\nu+\Gamma(\nu)\left(\frac{k}{2 a H}\right)^{-\nu}\right) \,.
\end{equation}
The power spectrum of $\chi$ during inflation is determined by
\begin{equation}
    \label{eq:powerspectrumgen}
    \mathcal{P}_{\chi} \; \equiv \; \frac{k^3}{2\pi^2} |\chi_k|^2 \, .
\end{equation}
Assuming that $\nu \in \mathbb{R}$, the mode function is approximated as~\cite{Liddle:1999pr, Mukhanov:1990me}
\begin{equation}
    |\chi_k(k \ll a H)| \; \simeq \; \frac{1}{\sqrt{2}H} \left(- \eta H \right)^{3/2} \left(\frac{k}{aH} \right)^{-\nu} \, ,
\end{equation}
and combining it with the power spectrum expression yields
\begin{equation}
        \mathcal{P}_{\chi} \; = \; \frac{H^2}{4\pi^2} \left(\frac{k}{a H} \right)^{3 - 2\nu} \, ,
\end{equation}
where $\nu$ is real. For minimal coupling $\xi = 0$ and massless limit $m_{\chi} = 0$, this results in the well-known scale-invariant power spectrum $\mathcal{P}_{\chi} = H^2/4\pi^2$. In the IR regime with $k \ll aH$, the long-wavelength modes freeze upon exiting the horizon, leading to no additional particle production upon their reentry. The particle occupation number then relates to the mode functions as $n_k(k, \eta) = |\beta_k|^2 \sim |\chi_k|^2$.

In this analysis, we concentrate on superheavy dark matter, characterized as such when $\nu$ becomes imaginary in Eq.~(\ref{eq:nuterm}), which occurs when $m_{\chi} \geq \frac{3}{2}H$ with $\xi = 0$.\footnote{For analysis of light scalar dark matter $m_{\chi} \lesssim H$ with nonminimal coupling, see~\cite{Garcia:2023qab}.} When $\nu \in \mathbb{R}$, the occupation number scales as $n_k \propto k^{-2 \nu}$, and the number density spectrum behaves as $\mathcal{N}_k \sim \mathcal{P}_{\chi} \propto k^{3-2\nu}$. In the minimal coupling scenario with $\xi = 0$, the occupation number is flat in the IR, with $n_k \propto k^0$, and $\mathcal{N}_k \sim \mathcal{P}_{\chi} \propto k^3$. For conformal coupling, with $\xi = 1/6$, $\nu = \sqrt{\frac{1}{4} - \frac{m_{\chi}^2}{H^2}}$, remaining real for $m_{\chi} \leq H/2$. To ensure that $\nu \in \mathbb{R}$, from Eq.~(\ref{eq:nuterm}), we find the condition 
\begin{equation}
    \label{eq:xiexp}
    \xi \leq \frac{3}{16} - \frac{1}{12} \frac{m_{\chi}^2}{H^2} \, .
\end{equation}
For $\xi < 0$, a red-tilted number density spectrum arises when $\xi < -\frac{1}{12} \frac{m_{\chi}^2}{H^2}$, leading to large IR divergence and high number density, discussed further in subsequent sections.

When $\nu$ becomes imaginary, the absolute value of the mode function simplifies to~\cite{Garcia:2023awt}
\begin{equation}
    |\chi_k (k \ll a H)| \; \simeq \;  \frac{1}{\sqrt{H}} \left(- \eta H \right)^{3/2} \frac{1}{\sqrt{2 \tilde{\nu}}} \, ,
\end{equation}
where $\tilde{\nu} = i \nu$ and $\tilde{\nu} \in \mathbb{R}$. The corresponding power spectrum then becomes~(\ref{eq:powerspectrumgen})
\begin{equation}
    \mathcal{P}_{\chi} \; \simeq \; \frac{1}{4\pi^2 a^3} \frac{k^3}{H \tilde{\nu}} \, .
\end{equation}
In this scenario, the occupation number is scale-invariant, with $n_k \propto k^0$, and the comoving number density spectrum scales as $\mathcal{N}_k \sim \mathcal{P}_{\chi} \propto k^3$. 
With increasing $\tilde{\nu}$, the power spectrum is further suppressed. The IR spectrum is exponentially suppressed as
$n_k \propto \exp(-c m_{\rm eff}/H)$, and the UV tail follows an exponential decay $\exp(-d k^2)$, where $c$ and $d$ are model-dependent constants~\cite{Chung:1998bt, Chung:2018ayg, Racco:2024aac}. Notably, as the nonminimal coupling $\xi > 0$ increases, it leads to a more suppressed number density. However, larger values of $\xi$ lead to efficient parametric resonance during reheating, resulting in significant gravitational particle production. The analytical mode solution for imaginary $\nu$ reveals  small oscillations in the IR due to interference effects~\cite{Kolb:2023dzp, Garcia:2023awt}, observable in our numerical solutions presented in subsequent sections (see Fig.~\ref{fig:nkvsk_negxi_1}).

\subsection{Steepest Descent Method}
\label{sec:steepestdescent}
For superheavy spectator fields, one can assume that the particle occupation number is small, with $|\beta_k|^2 \ll 1$. This assumption holds as long as the nonminimal coupling is not excessively large, with $12\xi H^2 \lesssim m_{\chi}^2$, which prevents significant resonant production during reheating.
Based on this assumption, the initial approximation for $\beta_k(\eta)$ up to leading order is given by~\cite{Kofman:1997yn}
\begin{equation}
\label{eq:bogbetainitial}
\beta_k(\eta) \simeq  \int_{\eta_i}^{\eta} d \eta' \,  \frac{\omega'_k}{2\omega_k} \exp\Big(-2i \Omega_k(\eta')\Big) \, ,\quad
\Omega_k(t) \equiv \int_{\eta_i}^{\eta} d \eta' \, \omega_k(\eta') \, .
\end{equation}
This approximation is valid provided that $\omega_k'/2\omega_k > 1$. We proceed to analyze the analytical computation of the particle occupation number $n_k = |\beta_k|^2$ using the steepest descent method~\cite{Chung:1998bt, Enomoto:2013mla, Enomoto:2020xlf, Racco:2024aac}. Here we briefly discuss this method and apply it for the superheavy mass case with nonminimal coupling (see the given references for a comprehensive study). This method focuses on the dominant contributions to the integral~(\ref{eq:bogbetainitial}) occurring near the poles in the complex $\eta$ plane. By adjusting the integration contour for $\beta_k$, we ensure that it passes through these poles along paths where the real part of the exponential term decreases most sharply, while the imaginary part remains constant. This approach ensures that the integrand rapidly diminishes as it moves away from the poles. The dominant contributions are computed by setting $\frac{\d}{\d \eta} \Omega_k (\eta) = 0$ and finding the poles, leading to
\begin{equation} 
   \omega_k (\eta_n) \; = \; 0 \,,
\end{equation} 
where $\eta_n$ represents the position of the poles. We keep our discussion general and consider the contribution arising from all $n$ poles. Following~\cite{Chung:1998bt, Racco:2024aac}, we deform the contour so that the poles lie in the lower half-plane. 

We first evaluate the contribution $\exp \left(-2i \Omega_k(\eta_n)\right)$. We expand the complex conformal time in terms of its real and imaginary components as $\eta_n = u_n + i v_n$ and decompose the integral as follows:
\begin{equation}
    \int_{\eta_i}^{\eta_n} d\eta' \omega_k(\eta') \; = \; \int_{\eta_i}^{u_n} d\eta' \omega_k(\eta') + \int_{u_n}^{\eta_n} d\eta' \omega_k(\eta') 
    \, .
    \label{eq:Omega_split}
\end{equation}
The first term is real and introduces a phase for each pole's contribution. Although this term generally cancels out when examining individual poles, it can influence the cumulative effect of multiple poles, despite the smaller contribution from the real axis integral. Consequently, we focus primarily on the contribution from the second, imaginary term. Given that $\omega_k(\eta_n)$ remains approximately constant from $u_n$ to $\eta_n$, we approximate the integral as
\begin{equation}
\label{eq:approxwkbintfull1}
\int_{\eta_i}^{\eta_n} \d \eta' \omega_k(\eta') \simeq i\, \Im \left( \int_{u_n}^{\eta_n} \d \eta' \omega_k(\eta') \right)
\; \simeq \; i \omega_k(u_n) v_n  \,,
\end{equation}
keeping only the dominant imaginary contribution. This approximation allows us to express the occupation number as
\begin{equation}
\label{eq:steepestdescapp1}
n_k \; = \; |\beta_k|^2 \; \simeq \; \frac{\pi^2}{9} 
\sum_n \exp\left( 4 \omega_k(u_n)v_n  \right)\, .
\end{equation}
This general expression accounts for the contribution from all $n$ poles. The computation of the prefactor $\pi^2/9$ and detailed analytical calculations are provided in Refs.~\cite{Chung:1998bt, Racco:2024aac}.

Next, we define the mode frequency~(\ref{eq:modefreq}) as
\begin{equation}
    \omega_k^2(\eta_n) \; = \; k^2 + a^2(\eta_n) m_{\rm eff}^2(\eta_n)  \; = \; k^2 + m_{\chi}^2 g(\eta_n) \; = \; 0 \,, \quad g(\eta_n) \; = \; -\frac{k^2}{m_{\chi}^2} \, ,
\end{equation}
where
\begin{equation}
     g(\eta) \; = \; \left(1 + \frac{\left(\frac{1}{6} - \xi \right)R(\eta)}{m_{\chi}^2} \right) a(\eta)^2\, .
\end{equation}
Taylor expanding $g(u_n)$, we find that the real part of the contribution up to quadratic order is given by
\begin{equation}
    g(u_n) - \frac{1}{2}v_n^2 g''(u_n) \; \simeq \; - \frac{k^2}{m_{\chi}^2} \, , 
\end{equation}
leading to
\begin{equation}
    \frac{1}{2}v_n^2 g''(u_n) \; = \; \frac{\omega_k^2(u_n)}{m_{\chi}^2}\, ,\qquad v_n \; = \; \sqrt{\frac{2}{g''(u_n)}} \frac{\omega_k(u_n)}{m_{\chi}}  \, .
\end{equation}
Combining these results with Eq.~(\ref{eq:steepestdescapp1}), the particle occupation number simplifies to
\begin{equation}
\label{eq:finalapprox}
|\beta_k|^2 \; \simeq \; \frac{\pi^2}{9} 
\sum_n \exp\left(\frac{4\sqrt{2}}{\sqrt{g''(u_n)}} \frac{k^2 + m_{\chi}^2 g(u_n)}{m_{\chi}} \right)\, .
\end{equation}
In this expression, all time-dependent quantities are evaluated at real values $\eta_n = u_n$. This equation reveals that for superheavy masses, the occupation number is exponentially suppressed. For more realistic inflation models such as Starobinsky or T-models, the pole contributions $g(u_n)$ and $g''(u_n)$ require fully numerical evaluation. However, we apply this functional form for our general fits in subsequent sections.

\section{Production Regimes}
\label{sec:production}
In this section, we explore the production of superheavy dark matter in scenarios with nonminimal coupling. We consider three cases: $\xi = 0$, $\xi < 0$, and $\xi > 0$. Building on findings from the previous section, we show that since the power spectrum is proportional to the square of the mode function, the comoving number density spectrum is proportional to $\mathcal{N}_k \sim \mathcal{P}_{\chi} \sim |\chi_k|^2$. The spectrum scales as
\begin{equation}
\begin{aligned}
    &\mathcal{N}_k \propto k^{3 - 2 \nu} \,, \qquad~&&\textrm{when~$\nu$~is~real} \, , 
    \\
    &\mathcal{N}_k \propto k^{3}\,, \qquad~&&\textrm{when~$\nu$~is~imaginary} \, , 
    \label{eq:comovingspectra}
\end{aligned}  
\end{equation}
for $k \ll a H$, where $\nu$ is given by Eq.~(\ref{eq:nuterm}). 

When $\xi \geq 0$ and $m_{\chi} > \frac{3}{2} H$, $\nu$ becomes imaginary, resulting in $\mathcal{N}_k \propto k^3$ and $n_k = |\beta_k|^2 \propto k^0$. This scaling holds for minimal coupling $\xi = 0$ and conformal coupling $\xi = 1/6$. We note that for $\xi \sim \mathcal{O}(1)$, the gravitational production during inflation is increasingly suppressed as $\xi$ increases. However, once $\xi$ becomes sufficiently large,
efficient gravitational production is observed during reheating due to parametric resonance, leading to significant post-inflationary production.

For $\xi < 0$ and $m_{\chi} > \frac{3}{2} H$, the parameter $\nu$ can be either real or imaginary, depending on the magnitude of $\xi$, and it remains real as long as the condition in Eq.~(\ref{eq:xiexp}) is satisfied. When $\nu > 3/2$, the comoving number density spectrum becomes red-tilted in the IR regime, leading to an IR divergence and substantial particle production. This divergence is regularized by introducing a natural IR cutoff, given by present-day comoving momentum, $k_0 = a_0 H_0$, under the assumption that this scale was inside the horizon at the beginning of inflation~\cite{Starobinsky:1994bd, Ling:2021zlj, Herring:2019hbe}. Due to this divergence, it is necessary to consider both isocurvature constraints and backreaction effects.

A qualitative summary of behavior across different $\xi$ regimes is illustrated in Fig.~\ref{fig:general}. We now proceed to analyze each scenario, both analytically and numerically, in greater detail.
\subsection{$\xi = 0$}
\label{sec:xizero}
In this section, we focus on the minimal production of superheavy spectator scalar field with $\xi = 0$. A comprehensive analytical and numerical treatment of this scenario has been recently explored in~\cite{Racco:2024aac}. First, we provide a detailed analytical study of the gravitational production of the scalar field in a de Sitter background using the steepest descent method. Second, we present numerical results and provide analytical fits across a range of masses for the Starobinsky model of inflation.

A crucial aspect of this analysis involves the definition of the adiabaticity parameter, which depends on the adiabatic basis and the choice of time. To resolve issues related to adiabaticity at late times, one could use conformal time while incorporating higher-order terms in the adiabatic expansion appearing in the mode frequency to resolve this issue~\cite{Corba:2022ugu}. To ensure that the mode frequency is adiabatic at late times, in a de Sitter background, we use cosmic time $t$ instead because conformal time does not behave adiabatically at late times for $\eta \rightarrow 0^{-}$ (see Appendix~\ref{app:cosmictime}).

For a de Sitter background, where the Ricci scalar is $R = -12H^2$, and assuming a constant Hubble parameter $H$, we use the mode equation~(\ref{eq:omega_k}) and set $\omega_k(t_n) = 0$:
\begin{equation}
    \frac{k^2}{e^{2H t}} +\mu^2 H^2 \; = \; 0 \, ,
\end{equation}
where 
\begin{equation}
    \mu^2 \; = \; \frac{m_{\chi}^2}{H^2} - \frac{9}{4} \, ,
\end{equation}
resulting in the condition
\begin{equation}
    H t_n \; = \; \log (\frac{k}{\mu H}) - i \frac{\pi}{2} + i n\pi \, , \quad n \in \mathbb Z \, .
\end{equation}

Focusing solely on the dominant pole contribution in the lower-half plane, we choose $n = 0$, as this is nearest to the real time axis. The particle occupation number then becomes
\begin{equation}
n_k \simeq  | \beta_k|^2 \simeq \frac{\pi^2}{9} \exp 
\left( -2 i \int_{t_n^*}^{t_n} dt' \, \omega_k(t') \right) \, .
\label{eq:nk_dS}
\end{equation}
The exponent is then:
\begin{equation}
\begin{aligned}
-2 i \int_{t_n^*}^{t_n} d t' \, \omega_k ( t') & \; = \; -2i \left(\mu H t - \frac{\sqrt{\frac{k^2}{e^{2Ht}}+\mu^2 H^2}}{H} +  \mu \log \left[\mu \left(H \mu + \sqrt{\frac{k^2}{e^{2Ht}}+\mu^2 H^2} \right) \right] \right) \bigg|_{t_n^*}^{t_n} \\
& \; = \;  \; - 2\pi \mu  \, .
\end{aligned}
\end{equation}
Thus, the resulting occupation number is given by
\begin{equation}
      n_k \; = \; |\beta_k|^2 \; \simeq \; \frac{\pi^2}{9} e^{-2 \pi \mu} \, . 
\end{equation}
This value agrees with findings from multiple studies~\cite{Anderson:2013ila, Anderson:2013zia, Markkanen:2016aes, Li:2019ves, Corba:2022ugu, Racco:2024aac}.

In contrast to the idealized de Sitter background, realistic models of inflation often exhibit a larger contribution from infrared modes. This enhancement typically arises during the transition from a quasi-de Sitter phase to a matter- or radiation-dominated era (or other backgrounds). A crucial factor in this enhancement is the sudden change in the equation of state parameter, $w_{\phi} = p_{\phi}/\rho_{\phi}$, which leads to a significant violation of the adiabaticity conditions.

It is important to note that short-wavelength (UV) modes, where $k > k_{\rm end}$, do not cross the horizon during the inflationary phase. Instead, the production of these modes primarily occurs during the 
transition to the matter-dominated era and throughout reheating. A detailed analytical and numerical examination of particle production within realistic inflationary frameworks, such as Starobinsky or T-models of inflation, is discussed in~\cite{Racco:2024aac}. Given the complexities of deriving the exact form of the scale factor $a(t)$ and locating its poles in inflation models, we compute the particle abundance entirely numerically for the Starobinsky model, with the key expressions and parameters given in Section~\ref{sec:inflation}. The results provided in this study are applicable to various inflation models.

For superheavy scalars with $\xi = 0$, it can be demonstrated that the post-inflationary gravitational particle production is primarily dominated by the initial oscillations of the inflaton. In inflationary models like the Starobinsky model~(\ref{eq:modestarobinsky}), which features a quadratic minimum, the energy density of the inflaton redshifts as matter, with $\rho_{\phi} \propto a^{-3}$, and the inflaton envelope redshifts as $\phi \propto a^{-3/2}\sim 1/t$. This decrease in amplitude causes the Hubble parameter to exhibit fluctuations, or `wiggles', which are proportional to $\sim H^2(t)/m_\phi^2$, arising from the non-smooth evolution over time~\cite{Racco:2024aac}.

We have numerically computed the occupation number $n_k = |\beta_k|^2$ and the comoving number density spectrum $\mathcal{N}_k/k_{\rm end}^3$ as a function of $k/k_{\rm end}$ evaluated at late time $a/a_{\rm end} = 100$, where $k_{\rm end} =  a_{\rm end} H_{\rm end}$ for a range of masses varying from $2H_I$ to $5H_I$. We note that at such late times, the spectra freeze and there is no additional particle production. In this case, the IR contribution is larger than that of pure de Sitter value of $n_k \simeq \exp(-2 \pi \mu)$, which is primarily due to the adiabaticity violation during the transition from a quasi-de Sitter state
($w_{\phi} = -1)$ to a matter-dominated background $(w_{\phi} = 0)$. Conversely, transition to a radiation-dominated background or other non-standard cosmological settings characterized by $w_{\phi} > 0$ tend to yield smaller gravitational particle production, as the violation of adiabaticity is less severe.

\begin{figure}[t!]
\centering
\includegraphics[width=0.81\textwidth]{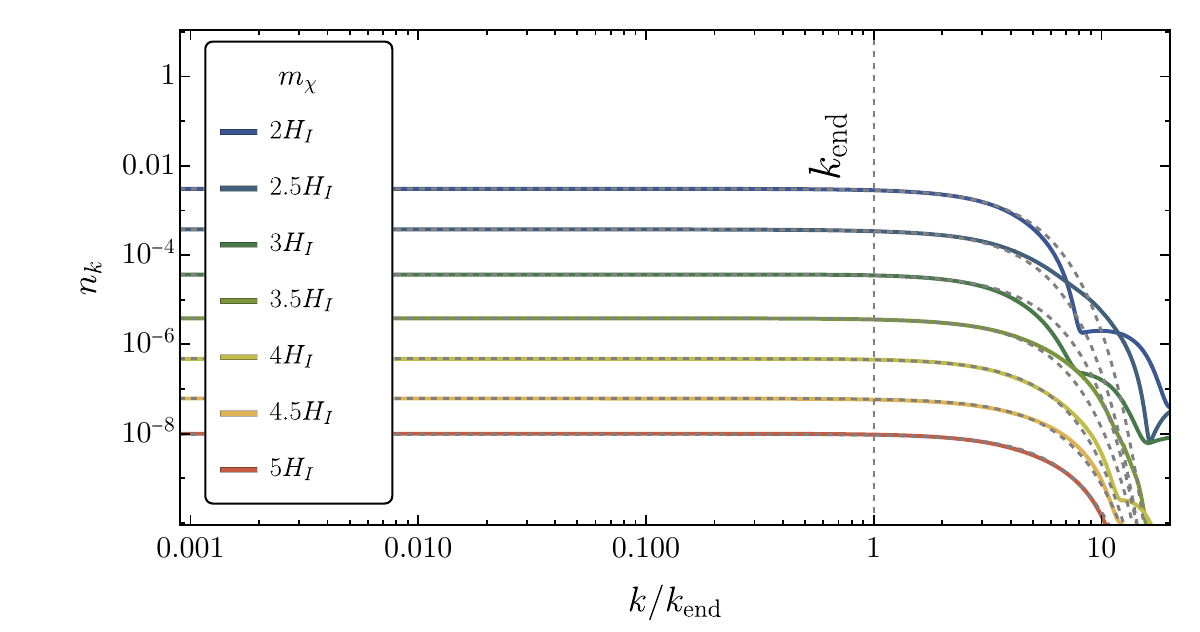}
\includegraphics[width=0.81\textwidth]{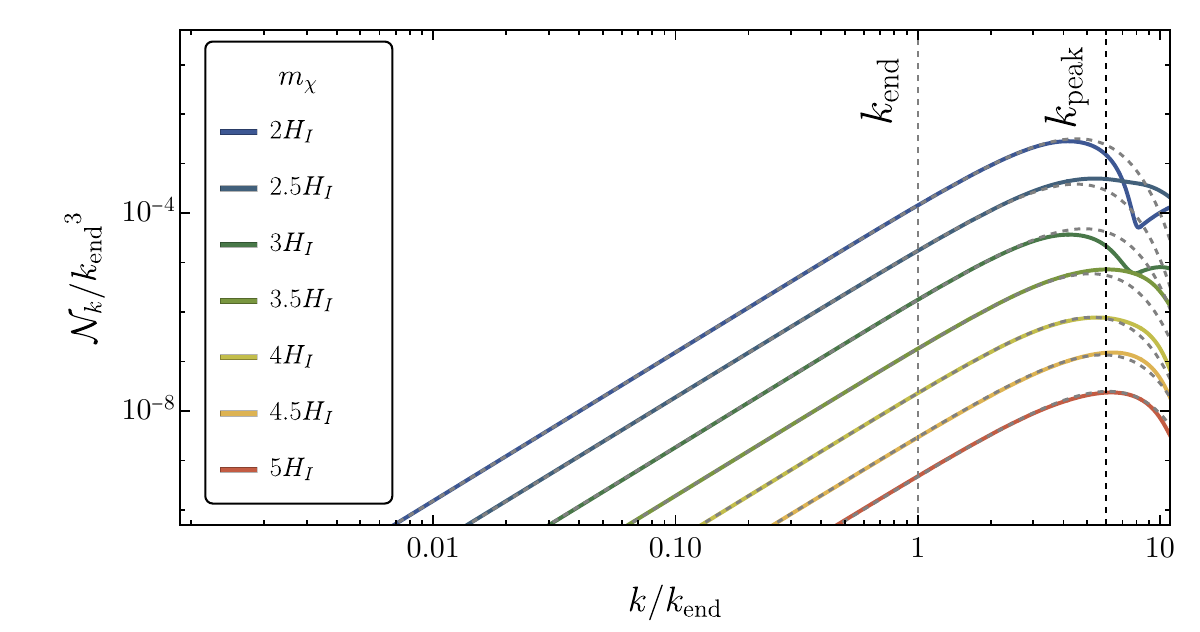}
\caption{Particle occupation number $n_k$ and the comoving number density spectrum $\mathcal{N}_k/k_{\rm end}^3$ as a function of $k/k_{\rm end}$ for a range of spectator field masses varying from $2H_I$ to $5H_I$. The spectra peak around $k/k_{\rm end} \sim 6$ for $m_{\chi} \geq 3.5H_I$, and the UV tail of the spectrum is exponentially suppressed.}
\label{fig:xi0plots}
\end{figure}

Our findings are summarized in Fig.~\ref{fig:xi0plots}. From the figure, one can see that the particle spectra peak around $k/k_{\rm end} \sim 6$ for masses $m_{\chi} \geq 3.5H_I$, with the spectra primarily dominated by the high-$k$ UV peak. This peak arises when the $k^2$ contribution becomes comparable to the effective mass term $a^2 m_{\rm eff}^2$ in the mode frequency~(\ref{eq:modefreq}). 

To account for the transition and subsequent particle production during the reheating phase, we use the analytical approximation given by Eq.~(\ref{eq:finalapprox}) and introduce the fit of the form:
\begin{equation}
   n_k \; = \; |\beta_k|^2 \; \simeq \; \frac{\pi^2}{9} \exp \left(-\frac{c(k/k_{\rm end})^2}{(m_{\chi}/H_I)} - \frac{d m_{\chi}}{H_I} \right) \, ,
\end{equation}
where $c$ and $d$ are fitting constants. We note that for spectator field masses exceeding $4.5H_I$, the fitting constants $c = 0.20$ and $d = 3.71$ do not change, indicating that in the superheavy regime ($m_{\chi} \gg H_I$), the fit remains consistent. We summarize the fitting results in Table~\ref{table:beta_k}.

\begin{table}[h!]
\centering
\setlength{\tabcolsep}{12pt} 
\renewcommand{\arraystretch}{1.2} 
\begin{tabular}{cccc}
\toprule
Mass & $c$ & $d$ & $n_k^{(\rm IR)}$ \\
\midrule
$2.0H_I$ & $0.14$ & $2.94$ & $3.0 \times 10^{-3}$ \\
$2.5H_I$ & $0.18$ & $3.20$ & $3.7 \times 10^{-4}$ \\
$3.0H_I$ & $0.19$ & $3.43$ & $3.7 \times 10^{-5}$ \\
$3.5H_I$ & $0.20$ & $3.60$ & $3.8 \times 10^{-6}$ \\
$4.0H_I$ & $0.22$ & $3.66$ & $4.8 \times 10^{-7}$ \\
$4.5H_I$ & $0.20$ & $3.71$ & $6.2 \times 10^{-8}$ \\
$5.0H_I$ & $0.20$ & $3.71$ & $9.6 \times 10^{-9}$ \\
\bottomrule
\end{tabular}
\caption{Table of the fitting parameters $c$, $d$, and the particle occupation number for low-$k$ $n_k^{(\rm IR)}$ for a range of masses.}
\label{table:beta_k}
\end{table}
It is important to note that while achieving very good fits for the low-$k$ infrared region is feasible, accurately fitting the high-$k$ ultraviolet peak and tail proves significantly more challenging, particularly for the lower mass ranges. This difficulty arises primarily because the masses $2H_I$, $2.5H_I$, and $3H_I$ are not yet in the regime $m_{\chi} \gg H_I$, leading to less precise fits at the peak. However, as the mass increases, the fits improve significantly in the UV, making it easier to delineate and accurately model both the IR plateau and the exponentially suppressed UV tail in $n_k$. We find excellent fits for masses $m_{\chi} \geq 4H_I$.

Next, we discuss the effects of nonminimal coupling.
\subsection{$\xi > 0$}
\label{sec:xipos}
We now explore the gravitational production of superheavy scalar fields with a positive nonminimal coupling, $\xi > 0$. In this case, as $\xi$ increases, particle production during inflation tends to decrease significantly due to the increasing effective mass $m_{\rm eff}^2 = m_{\chi}^2 -2H^2 + 12 \xi H^2$. For minimal coupling ($\xi = 0$), the effective mass simplifies to $m_{\rm eff}^2 = m_{\chi}^2 -2H^2$, and for conformal coupling ($\xi = 1/6$), the effective mass is $m_{\rm eff}^2 = m_{\chi}^2$. 

As $\xi$ increases, particle production becomes more efficient primarily due to effective parametric resonance, which leads to substantial post-inflationary particle production.\footnote{This effect mirrors the impact of introducing a coupling $\frac{1}{2}\sigma \phi^2 \chi^2$ in the Lagrangian, which leads to the effective mass of $m_{\rm eff}^2 = m_{\chi}^2 + \sigma \phi^2 + \frac{1}{6}R$. With increasing $\sigma$, similar to increasing $\xi$, the model can successfully avoid isocurvature constraints, and higher values of $\sigma$ result in a broad parametric resonance, leading to efficient post-inflationary particle production~\cite{Garcia:2022vwm, Garcia:2023awt}.} We now study the specifics of this production regime.

\begin{figure}[h!]
\centering
\includegraphics[width=0.81\textwidth]{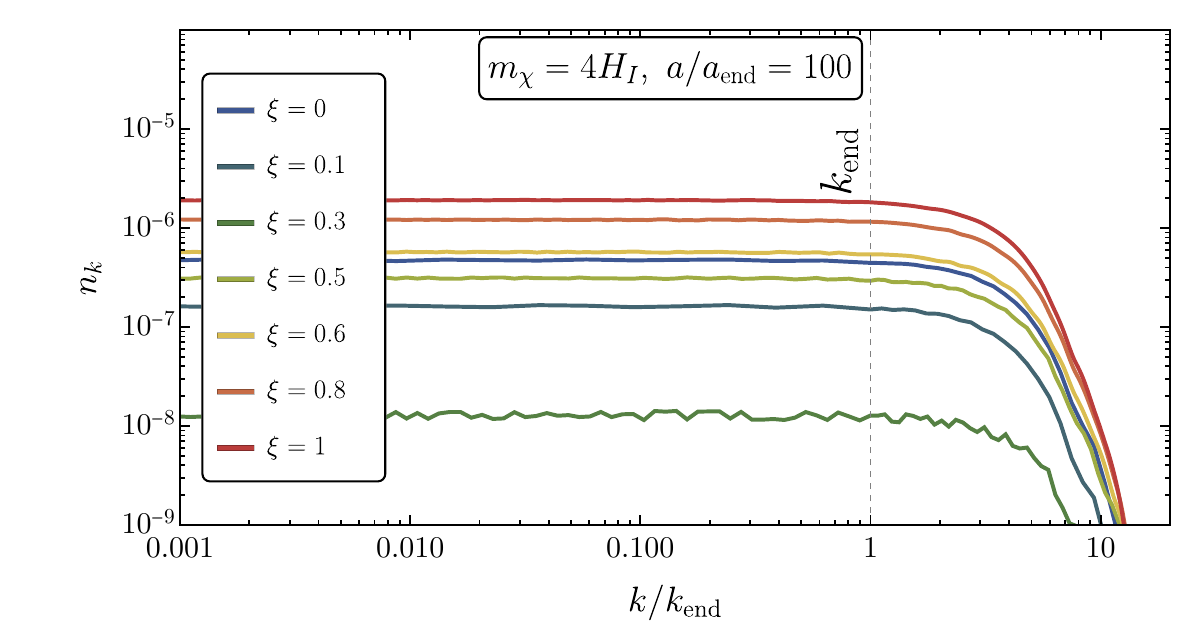}
\includegraphics[width=0.81\textwidth]{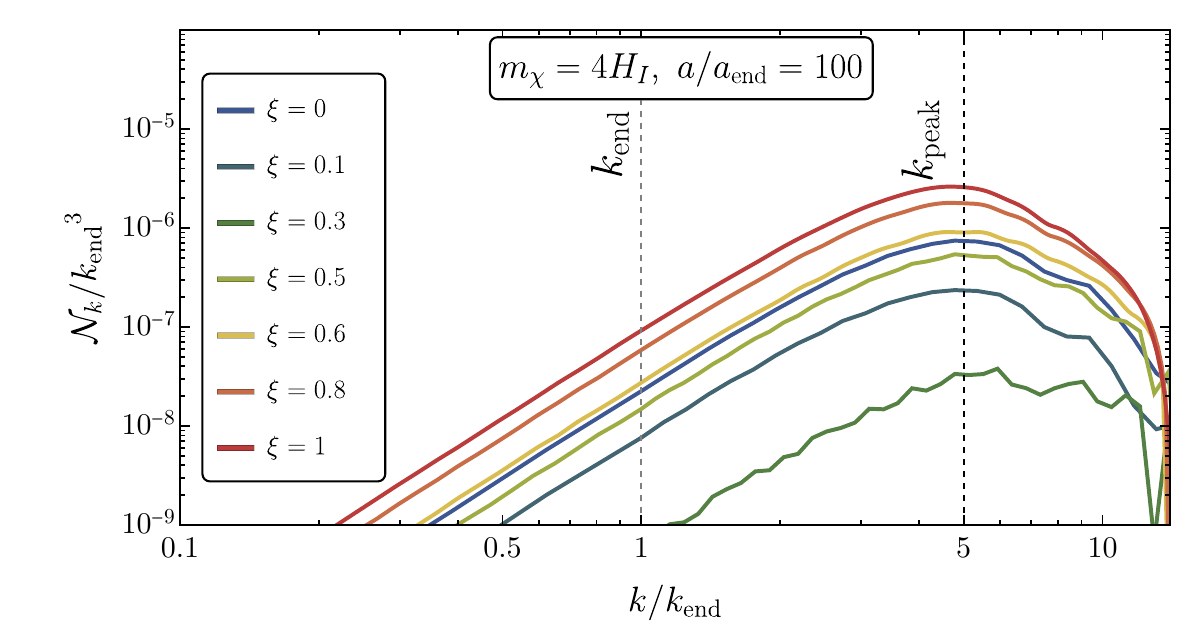}
\caption{Particle occupation number $n_k$ and the comoving number density spectrum $\mathcal{N}_k/k_{\rm end}^3$ as a function of wavenumber normalized to the end of inflation $k/k_{\rm end}$,for a range of nonminimal couplings $0 < \xi \leq 1$. The spectra peak around $k/k_{\rm end} \sim 5$, and the overall effect of $\xi$ on the position of this peak is negligible. However, as $\xi$ increases up to $\xi = 0.3$,  the spectra not only become more suppressed but also exhibit increased noise. 
The spectra then continues growing for $\xi > 0.3$. As before, the high-$k$ ultraviolet (UV) tail of the spectrum remains exponentially suppressed.}
\label{fig:xiposplots1}
\end{figure}

\subsubsection{Small Coupling $0 < \xi \leq 1$}
For nonminimal couplings $\xi \leq 1$, a notable trend emerges: as $\xi$ increases, the infrared contribution to particle production becomes more suppressed, leading to less efficient post-inflationary. This inefficient production arises from the smallness of $\xi$, which is insufficient to trigger significant parametric resonance. We observe that the particle occupation number density $n_k$ becomes increasingly suppressed and noisier up to $\xi = 0.3$ as the effective mass increases, largely due to interference effects. For a choice of $m_{\chi} = 4H_I$, the comoving number density spectra peak around $k_{\rm peak} \sim 5$, with small nonminimal coupling not altering this position of the peak. Numerically, we find that for $\xi = 0.3$, $n_k^{(\rm IR)} \simeq 1.4 \times 10^{-8}$, which is significantly more suppressed than $n_k^{(\rm IR)} \simeq 4.8 \times 10^{-7}$ for $\xi = 0$, indicating decreased efficiency in dark matter production. However, as $\xi$ increases further, $n_k$ begins to grow due to more efficient particle production during the transition from inflation to a matter-dominated universe and reheating. At $\xi = 0.6$, 
$n_k^{(\rm IR)} \simeq 5.7 \times 10^{-7}$, close to the minimally coupled scenario.

Analytically fitting these distributions proves challenging, unlike the case with $\xi = 0$, due to the increased effective mass of the spectator field and the enhanced parametric resonance during reheating. Our numerical results for small nonminimal couplings are summarized in Fig.~\ref{fig:xiposplots1}.
\subsubsection{Large Coupling $\xi > 1$}

\begin{figure}[h!]
\centering
\includegraphics[width=0.81\textwidth]{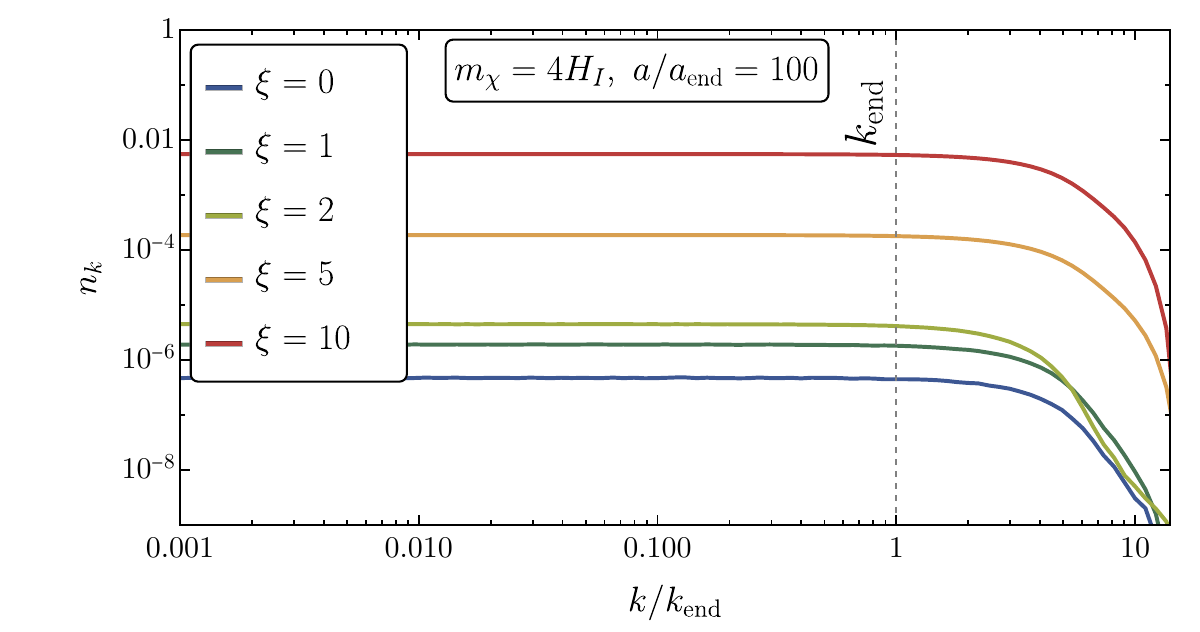}
\includegraphics[width=0.81\textwidth]{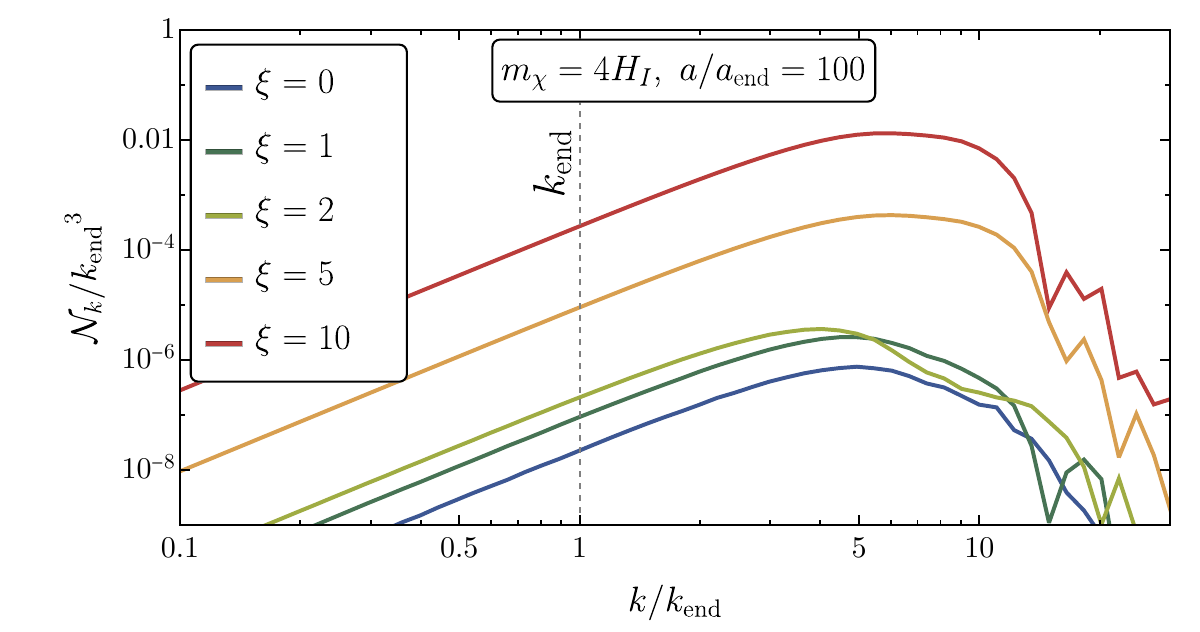}
\caption{Particle occupation number $n_k$ and the comoving number density spectrum $\mathcal{N}_k/k_{\rm end}^3$ as a function of $k/k_{\rm end}$ for a range of large nonminimal couplings $0 < \xi \leq 10$. As $\xi$ increases, the comoving spectra grow, driven by efficient parametric resonance. For $\xi = 0$, the spectrum peaks at $k/k_{\rm end} \sim 5$, but with increasing $\xi$, this peak progressively shifts to larger values. The highly efficient parametric resonance during the post-inflationary era significantly influences the behavior of the spectra, 
populating the UV tail and broadening the spectrum as $\xi$ grows. Moreover, the UV portion of the spectra becomes increasingly noisier due to interference effects. We note that the exponential suppression of the UV tail occurs only at higher values of $k/k_{\rm end}$
as $\xi$ increases.}
\label{fig:xiposplots2}
\end{figure}

For large nonminimal couplings $\xi \geq 1$, gravitational particle production differs both during and after inflation. During inflation, the effective mass of a spectator field is considerably increased, with $m_{\rm eff}^2 \simeq 12 \xi H^2$, which decreases the efficiency of particle production. However, post-inflationary production becomes significantly more efficient due to parametric resonance, resulting in a substantial increase in total abundance at late times. 

As $\xi$ increases, the UV peak of $\mathcal{N}_k$ begins to shift to higher values of $k/k_{\rm end}$, moving from $k/k_{\rm end} \sim 5$ when $\xi = 0$ to $k/k_{\rm end} \sim 8$ when $\xi = 10$. Importantly, the amplitude of the spectrum grows significantly, with the UV tail becoming increasingly populated due to particle production during reheating. Consequently, the spectra broaden, and the UV tail becomes noisy due to interference effects. For $\xi = 5$ and $10$, the UV tail becomes more populated, and exponential suppression becomes important at 
larger values of $k$. This behavior can be intuitively understood through the mode frequency equation~(\ref{eq:modefreq}). For large values of $\xi$, during reheating, the oscillating term $-\xi R(\eta)$ becomes more significant as $\xi$ increases, causing the $k^2$ term to dominate only at higher values of $k/k_{\rm end}$. Our numerical findings for a mass $m_{\chi} = 4H_I$ are summarized in Fig.~\ref{fig:xiposplots2}, with spectra evaluated at $a/a_{\rm end} = 100$.

We now proceed to discuss the analytical treatment of parametric resonance, explaining why post-inflationary gravitational particle production becomes extremely efficient for large values of $\xi$.
\subsubsection{Parametric Resonance}
\label{sec:parametricresonance}
In scenarios where the nonminimal coupling $\xi \gg 1$, the spectator field becomes strongly coupled to the oscillating background. For this analysis, we use cosmic time to describe the parametric resonance.

During inflaton oscillations around its minimum after inflation, the Ricci scalar can be approximated by
\begin{equation}
    \label{eq:riccigeneral}
    R \; \simeq \; -\dfrac{1}{M_P^2}\big(4V(\phi)-\dot{\phi}^2\big) \,,
\end{equation}
assuming the energy density of the spectator scalar field is subdominant compared to that of the inflaton energy density. At the end of inflation, the inflaton field, $\phi$, behaves as
\begin{equation}
    \phi(t) \; \simeq \; \phi_{\rm end} \left( \dfrac{a(t)}{a_{\rm end}} \right)^{-3/2} \cos\big( m_\phi(t-t_{\rm end}) \big) \,. 
\end{equation}
Using this expression together with Eq.~(\ref{eq:riccigeneral}), we find
\begin{equation}
    R \; \simeq \; - \dfrac{m_\phi^2 \phi_\text{end}^2}{2 M_P^2} \left( \dfrac{a(t)}{a_\text{end}} \right)^{-3} \Big( 1 + 3 \cos\big( 2 m_\phi(t-t_\text{end}) \big) \Big) \,,
    \label{eq:ricciapproxnew}
\end{equation}
where we neglected the rapidly decreasing terms proportional to the powers of $H/m_\phi$. 

The equation of motion for the rescaled field $\Psi_k \equiv  a^{3/2} \chi_k$ takes the form of the Mathieu equation:
\begin{equation}
    \dfrac{\diff^2 \Psi_k}{\diff z^2}  + \big[ A_k+2 \kappa \cos(4 z) \big]  \Psi_k \; = \; 0 \,,
\end{equation}
where $z \equiv m_\phi(t-t_\text{end})/2$, and 
\begin{equation}
    A_k \, \equiv \, \dfrac{4 k^2}{a^2 m_\phi^2} +  \dfrac{2  \xi\phi_\text{end}^2}{M_P^2}  \left( \dfrac{a_\text{end}}{a} \right)^3  , \qquad \kappa \equiv \dfrac{3 \xi \phi_\text{end}^2}{M_P^2} \left( \dfrac{a_\text{end}}{a} \right)^3\,,
    \label{eq:coeffsMathieu}
\end{equation}
in agreement with Refs.~\cite{Lebedev:2022vwf, Garcia:2023qab}. The Mathieu equation introduces parametric instabilities, potentially leading to parametric resonances when $\kappa \gg 1$, which translates to $\xi \gg 1$. As shown in Fig.~\ref{fig:xiposplots2}, with increasing $\xi$, the UV tail of the spectrum becomes significantly more populated, broadening the spectrum and elevating the peak considerably, particularly when $\xi \geq 5$.

The coefficients in the Mathieu equation~(\ref{eq:coeffsMathieu}) have a strong dependence on the scale factor, resulting in the exponential amplification of mode functions typically ceasing soon after reheating begins, around $a/a_{\rm end} \simeq 10$, equivalent to $\sim 15$ inflaton oscillations. However, with large values of $\xi$, the backreaction on the inflaton could lead to fragmentation of the condensate. Additionally, if the energy density of the spectator field becomes significant, the background value of the Ricci scalar could deviate from the estimate of Eq.~(\ref{eq:ricciapproxnew}). In such scenarios, our current analytical approach becomes inadequate, requiring full simulations of the coupled inflaton and spectator field system on a lattice, which we discuss in the subsequent section.

For light dark matter $m_{\chi} < H_I$, lattice results discussed in ~\cite{Figueroa:2021iwm, Lebedev:2022vwf} suggest that the backreaction becomes significant when the energy density of the spectator field does not exceed $1\%$ of the inflaton energy density at the end of inflation. Numerically, we find that for light scalar fields $\xi \lesssim 70$ for $\rho_\chi/\rho_\text{end}<0.01$ and $\xi \lesssim 85$ for $\rho_\chi/\rho_\text{end}<0.1$, in agreement with Ref.~\cite{Lebedev:2022vwf}. In contrast, for superheavy spectator fields, the thresholds for significant backreaction are substantially higher, and in most cases, these values are excluded by constraints on dark matter abundance. We discuss these aspects in the following section.

\subsubsection{Backreaction and Lattice Results}
\label{sec:backposxi}
Here we present some examples where backreaction becomes significant and apply lattice simulations. We have used a modified {\tt ${\mathcal C}$osmo${\mathcal L}$attice} code~\cite{Figueroa:2020rrl, Figueroa:2021yhd} that includes nonminimal coupling to gravity.\footnote{I extend my gratitude to the authors of {\tt ${\mathcal C}$osmo${\mathcal L}$attice} for sharing the code with nonminimal couplings.} For these simulations, we employed a lattice with $N = 256$ sites/dimensions, exploring a spectrum of masses ranging from from $1.5H_I$ to $5H_I$ and a broad range of nonminimal coupling up to $\xi \sim 1000$.

Using the stress-energy tensor~(\ref{eq:tmunuscalar}), the equation of motion in the Jordan frame is given by~\cite{Figueroa:2021iwm, Lebedev:2022vwf}
\begin{equation}
    \rho_{\chi} \; = \; \langle T_{00}^{\chi} \rangle \; = \; \frac{1}{2a^2} \langle \chi'^2 \rangle + \frac{1}{2a^2} \langle (\nabla \chi)^2 \rangle + \langle  V(\chi) \rangle + 3 \xi H^2 \langle \chi^2 \rangle + \frac{6 \xi}{a} H \langle \chi \chi' \rangle - \frac{\xi}{a^2} \langle \nabla^2 \chi^2 \rangle \, .
\end{equation}

In models with nonminimal coupling, accurately interpreting the spectator field's energy density is complicated due to the last three terms in the stress-energy tensor. Although the spatial derivative term $\nabla^2 \chi^2$ 
can generally be disregarded as it averages to zero over space, other terms become critical, particularly for large $\xi$. The term $\frac{6 \xi}{a} H \langle \chi \chi' \rangle$ can grow substantially large and negative, potentially leading to a negative total energy density for the spectator field.

To account for the negative contributions to energy density and the kinetic mixing between the metric and the scalar field, we rely on lattice results for large $\xi$ values, where backreaction effects become significant. However, as we approach higher $\xi$ values, backreaction constraints together with dark matter abundance limits, discussed in Section~\ref{sec:dmabundance}, indicate that such high values of $\xi$ would be excluded by dark matter overproduction. We include the lattice studies for completion. For a more detailed discussion, see Refs.~\cite{Figueroa:2021iwm, Lebedev:2022vwf}.

We summarize the values of $\xi$ in Table~\ref{table:xicutoff} when the backreaction becomes important for a range of masses ranging from $1.5H_I$ to $5H_I$.\footnote{Introducing a self-interaction term for the spectator field, $\frac{1}{4}\lambda_{\chi} \chi^4$, would alter the backreaction limits, affecting the limits at which these effects become significant.}

\begin{figure}[h!]
\centering
\includegraphics[width=0.49\textwidth]{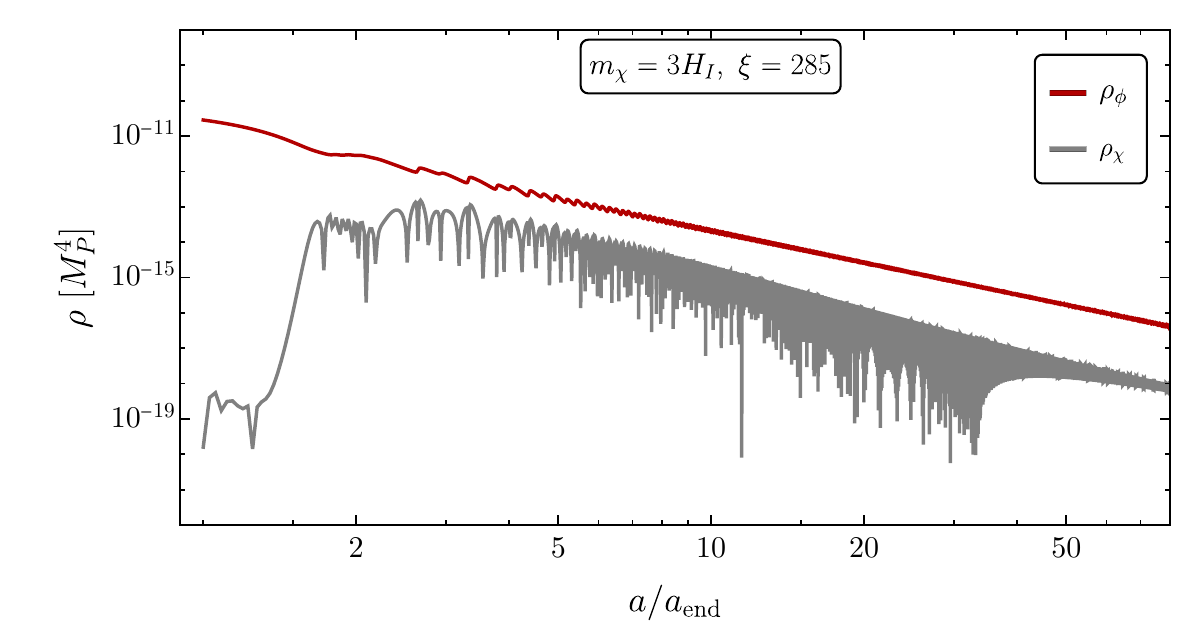}
\includegraphics[width=0.49\textwidth]{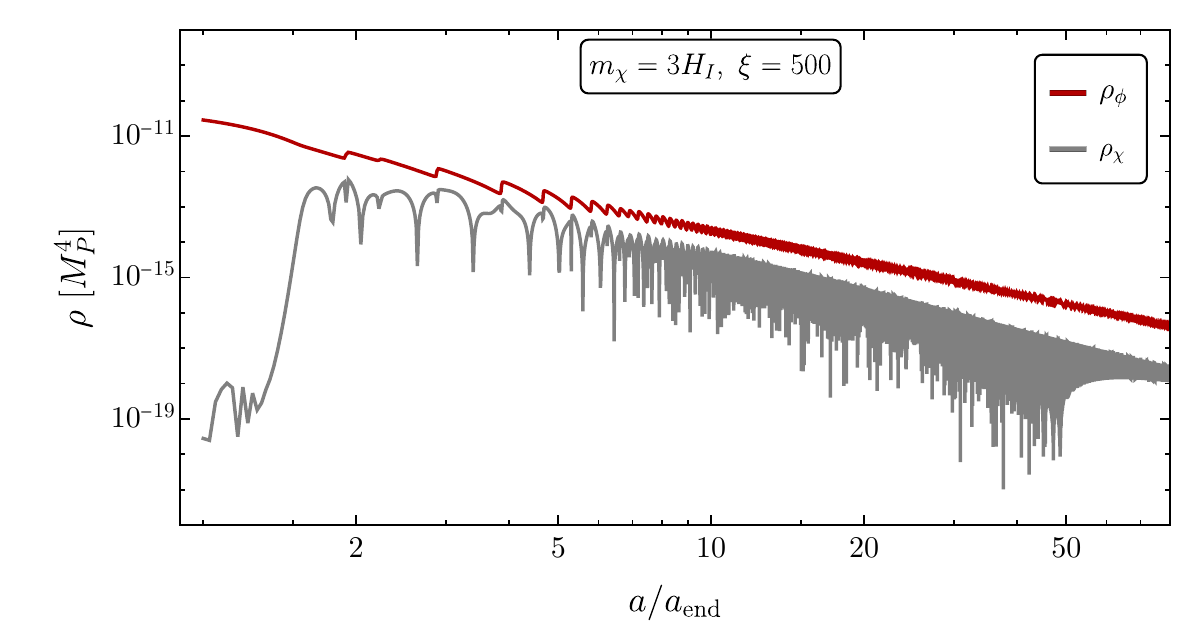}
\includegraphics[width=0.49\textwidth]{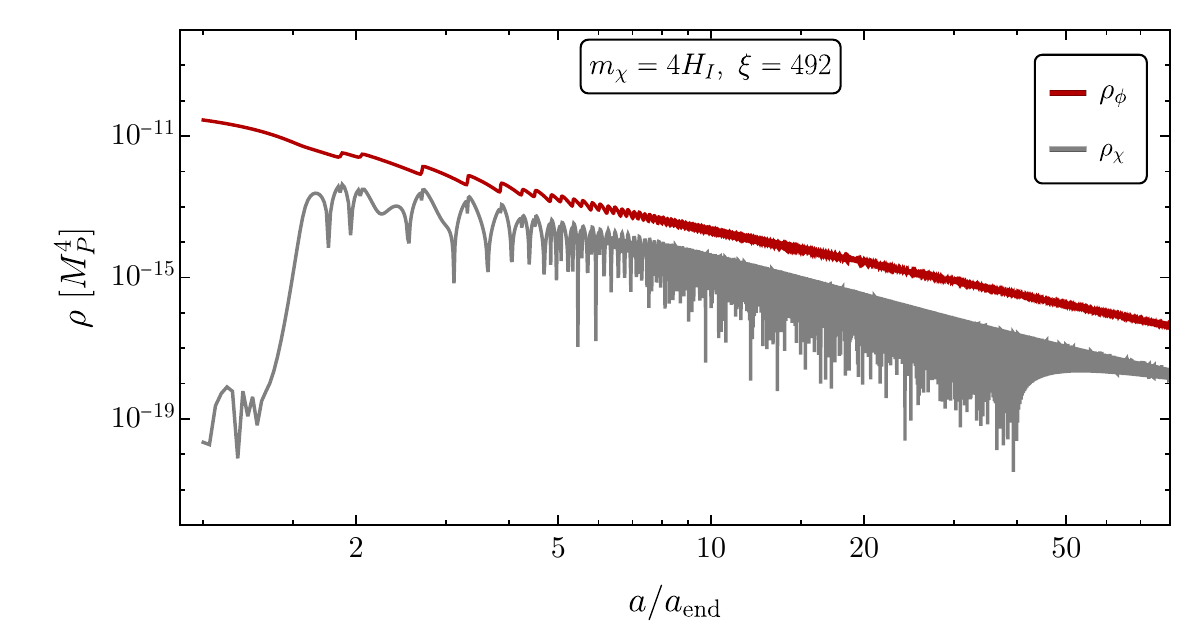}
\includegraphics[width=0.49\textwidth]{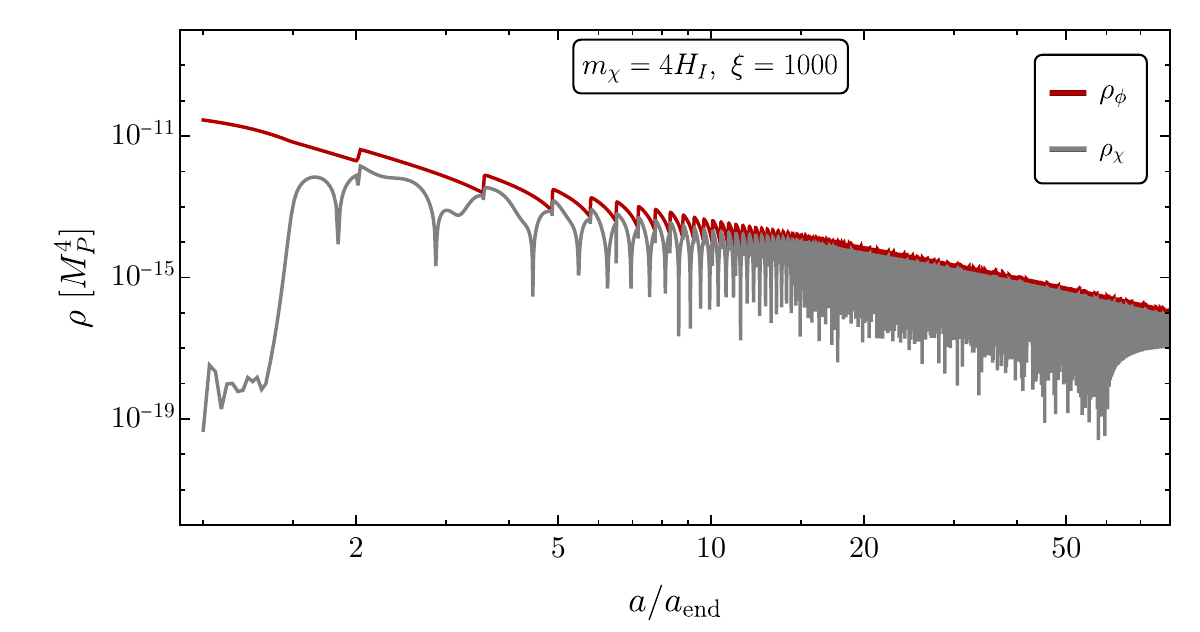}
\caption{The energy density as a function of the scale factor $a/a_{\rm end}$ in the backreaction regime, for a mass of $3H_I$ and nonminimal coupling $\xi = 285$ (top left panel) and $\xi = 500$ (top right panel), and for a mass of $4H_I$ and nonminimal coupling $\xi = 492$ (bottom left panel) and $\xi = 1000$ (bottom right panel). These results are obtained using a modified {\tt ${\mathcal C}$osmo${\mathcal L}$attice} code with nonminimal coupling with $N = 256$ sites/dimensions.}
\label{fig:xilatticeplots}
\end{figure}

Furthermore, lattice results illustrating backreaction effects for the Starobinsky model of inflation, with $\rho_{\rm end} = 3H_{\rm end}^2 M_P^2 \simeq 2.9 \times 10^{-11} M_P^4$, are displayed in Fig.~\ref{fig:xilatticeplots}, where we track the evolution of the energy densities of the inflaton, $\rho_{\phi}$, (red curve) and the spectator scalar field, $\rho_{\chi}$ (gray curve), as functions of $a/a_{\rm end}$. 
These results show spectator scalar masses of $m_{\chi} = 3H_I$ and $\xi = 285, 500$, and $m_{\chi} = 4H_I$ and $\xi = 492,1000$.

\begin{table}[h!]
\centering
\setlength{\tabcolsep}{12pt} 
\renewcommand{\arraystretch}{1.2} 
\begin{tabular}{ccccccccc}
\toprule
Mass & $1.5H_I$ & $2H_I$ & $2.5H_I$ & $3H_I$ & $3.5H_I$ & $4H_I$ & $4.5H_I$  & $5H_I$\\
\midrule
$\xi$ & $148$ & $245$ & $258$ & $285$ & $385$ & $492$ & $502$ & $525$ \\
\bottomrule
\end{tabular}
\caption{The values of $\xi$ at which the backreaction at which backreaction effects become important for masses ranging from $1.5H_I$ to $5H_I$.}
\label{table:xicutoff}
\end{table}

\subsection{$\xi < 0$}
\label{sec:xineg}
We now explore the effects of a negative nonminimal coupling $\xi < 0$ on the gravitational production of superheavy spectator fields. The negative $\xi$ influences the production in two significant ways: 1) Negative $\xi$ alters the parameter $\nu$, given by Eq.~(\ref{eq:nuterm}), during the inflationary period. If $\xi$ is significantly negative, this change can result in a red-tilted comoving number density spectrum. 2) A substantially negative $\xi$ enhances particle abundance, which can lead to a large amplitude of the isocurvature power spectrum and strong backreaction effects.
Additionally, in cases where $\xi \ll -1$, there is significant post-inflationary particle production due to very efficient parametric resonance, 
similar to the case with large $\xi \gg 1$ that we discussed previously.

To study the IR divergence when $\xi$ is negative, we first discuss the tachyonic mode growth. When the induced Hubble mass during inflation $12 \xi H^2$ is sufficiently large, it dominates over the bare mass term $m_{\chi}^2$ in the mode frequency $m_{\rm eff}^2 = m_{\chi}^2 + 12 \xi H^2$. In this case, the effective mass and the mode frequency squared~(\ref{eq:modefreq}) becomes negative, leading to tachyonic excitation of the corresponding mode. To qualitatively understand this effect, we use the second Friedmann equation
\begin{equation}
    R \; = \; - \frac{6a''}{a^3} \; = \; -\frac{1}{M_P^2} \left(4V - \frac{\phi'^2}{a^2} \right) \, .
\end{equation}
At the end of inflation, as discussed in Section~(\ref{sec:inflation}), $V(\phi_{\rm end}) = \phi'^{2}_{\rm end}/a_{\rm end}^2$ and $H_{\rm end}^2 M_P^2 = V_{\rm end}/2$, and the mode frequency can be expressed as
\begin{equation}
    \omega_k^2(\eta_{\rm end}) \; = \; k^2 + a_{\rm end}^2 \left(m_{\chi}^2 - H_{\rm end}^2 \left(1 - 6\xi \right) \right) \, .
\end{equation}
For the spectator field with mass $m_{\chi}$, if $m_{\chi}^2 < H_{\rm end}^2 \left(1 - 6\xi \right)$, the momentum modes $k^2 < a_{\rm end}^2(H_{\rm end}^2(1-6\xi) - m_{\chi}^2)$ will undergo tachyonic growth during inflation. 
For an accurate analysis of this phenomenon, it is important that these modes are well within the horizon at the beginning of inflation, and their evolution must be carefully tracked from very early times to accurately capture the dynamics. The extent of tachyonic growth for each mode is determined by the duration these modes spend outside the horizon, and how long their mode frequency $\omega_k$ remains imaginary. Moreover, for a largely negative nonminimal coupling $\xi \ll -1$, the comoving number density spectra become strongly red-tilted for $k \ll a H$ modes.

For $k \ll k_{\rm end}$, the comoving particle density spectrum scales as $\mathcal{N}_k \propto k^{3-2\nu}$ (See Eq.~(\ref{eq:comovingspectra}) and the discussion below it). This relationship leads to logarithmic divergence when $\nu = 3/2$, or $12 \xi = -m_{\chi}^2/H^2$. As $\xi$ becomes increasingly negative, the tachyonic growth of long-wavelength modes results in a larger divergence in the IR spectrum for $k \ll k_{\rm end}$. Such divergence requires a natural cutoff.

Regardless of the background renormalization due to these long-wavelength IR modes, it is always possible to introduce a natural cutoff for the comoving number density spectrum. These modes, which were superhorizon at the onset of cosmic inflation, only grow due to tachyonic instability during inflation. As suggested in~\cite{Starobinsky:1994bd, Ling:2021zlj, Herring:2019hbe}, this cutoff can be associated with the present-day comoving horizon scale, $k_0 = a_0 H_0$, under the assumption that this mode was within the horizon at the beginning of inflation. Importantly, the smaller wavenumber modes, which are outside our cosmological horizon, are counted as part of the homogeneous background. Furthermore, as the IR divergence becomes larger, the backreaction effects become significant. These effects and their implications are discussed further in the subsequent sections.

Despite the strong red-tilt in distributions when $\xi$ is largely negative, the behavior of short-wavelength UV modes remains similar to the case with $\xi = 0$. The peak of these distributions typically occurs between $k_{\rm peak} \sim 5-10 \, k_{\rm end}$, and the peak grows during reheating due to parametric resonance for large values of negative $\xi$. However, these distributions are dominated by the IR contribution when $\nu > 3/2$. As $\xi$ becomes increasingly negative, there is a shift in the peak towards larger values of $k/k_{\rm end}$, but the UV tail of the distribution remains strongly exponentially suppressed.
\subsubsection{Small Coupling $-1 \leq \xi < 0$}

\begin{figure}[h!]
\centering
\includegraphics[width=0.8\textwidth]{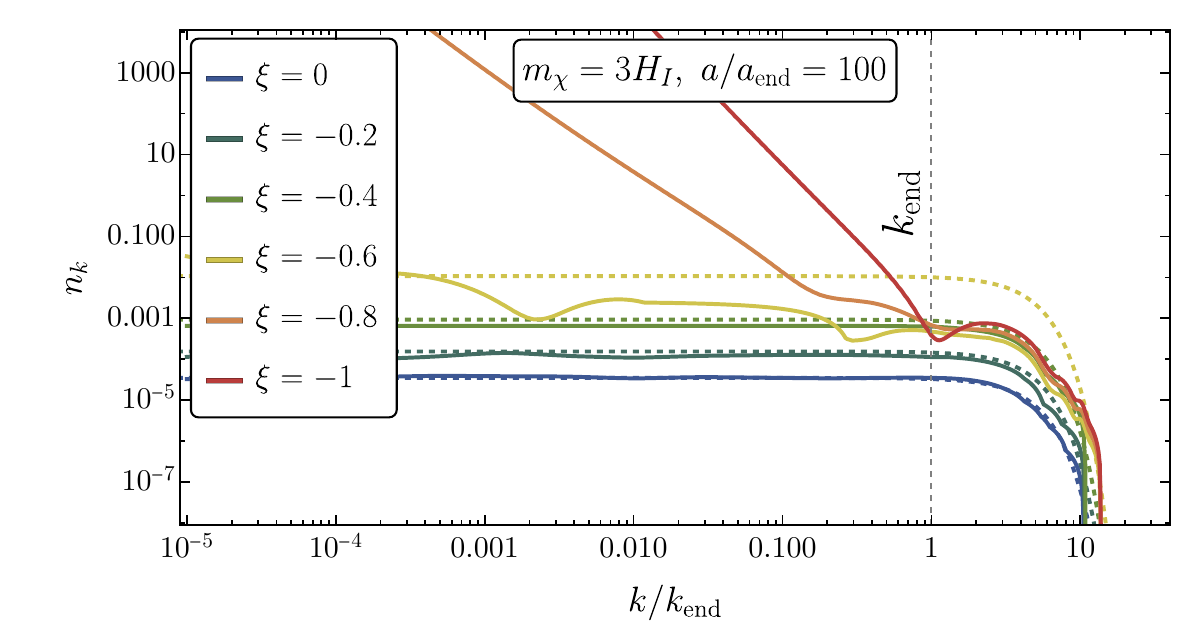}
\includegraphics[width=0.8\textwidth]{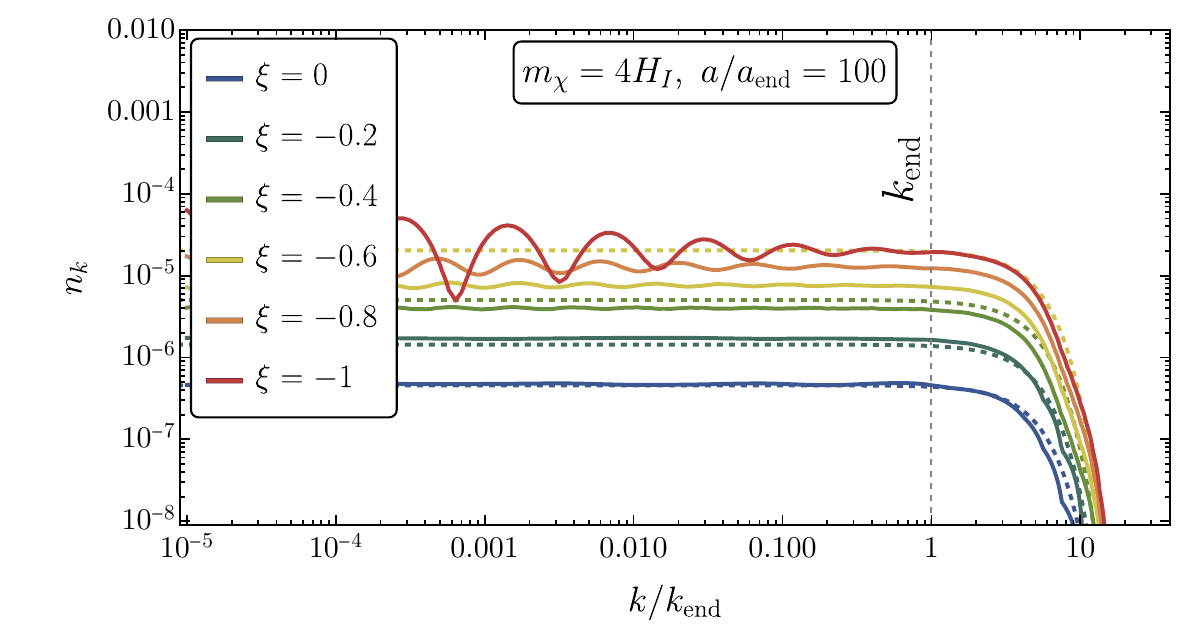}
\caption{Particle occupation number $n_k$ as a function of $k/k_{\rm end}$ for a range of small negative nonminimal couplings $-1 \leq \xi \leq 0$ for masses $m_{\chi} = 3H_I$ (top panel) and $m_{\chi} = 4H_I$ (bottom panel). As $\xi$ decreases, $n_k$ grows, and interference effects become more significant. For the mass of $m_{\chi} = 3H_I$, the spectrum shows a tilt in the IR region as $\xi$ decreases to $\xi \leq -0.6$.}
\label{fig:nkvsk_negxi_1}
\end{figure}

\begin{figure}[h!]
\centering
\includegraphics[width=0.8\textwidth]{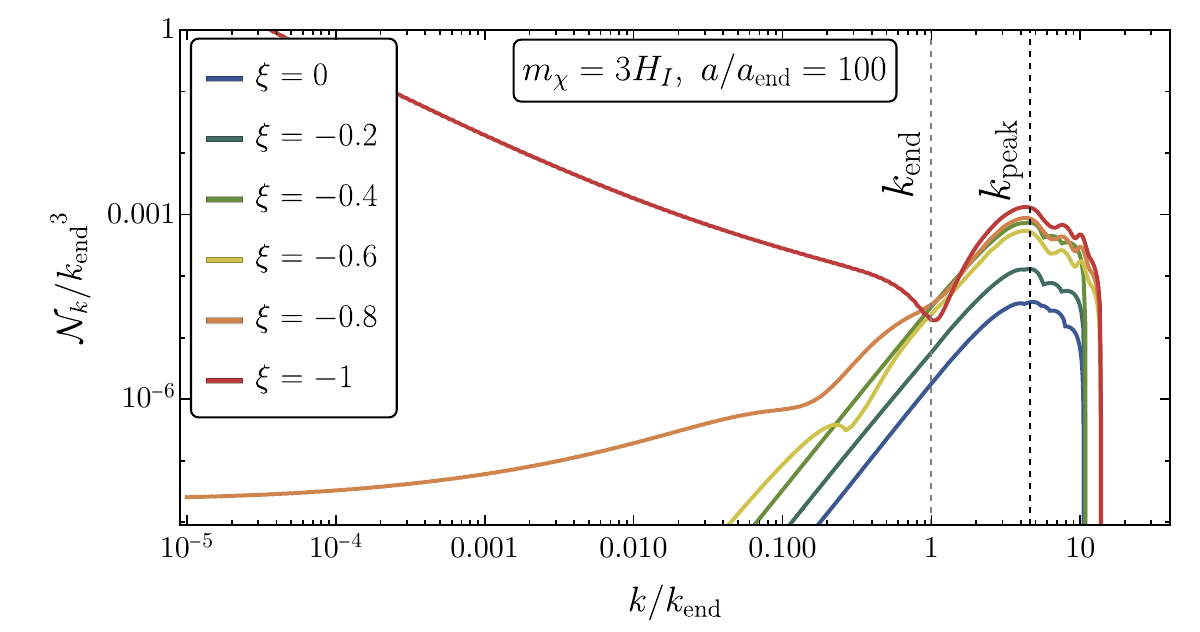}
\includegraphics[width=0.8\textwidth]{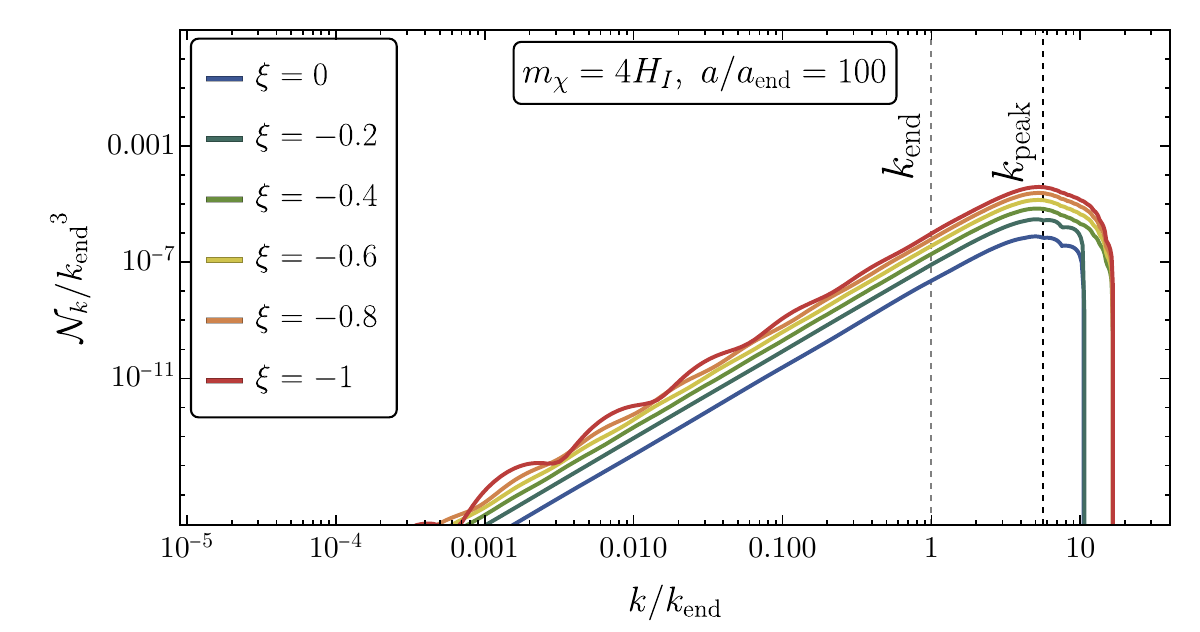}
\caption{Same as Fig.~\ref{fig:nkvsk_negxi_1}, but here we plot the comoving number density density spectrum $\mathcal{N}_k/k_{\rm end}^3$ as a function of $k/k_{\rm end}$. We observe that the spectra peak around $k/k_{\rm end} \sim 6$. As $\xi$ decreases, the distribution in the UV region broadens. As for the cases with $\xi \geq 0$, the UV tail remains exponentially suppressed.}
\label{fig:nkvsk_negxi_2}
\end{figure}

We now focus on scenarios where nonminimal coupling is relatively small, with $-1 \leq \xi < 0$. In these cases, the influence of nonminimal coupling is small and does not substantially enhance particle production during reheating.
For coupling values between $-0.5 \leq \xi < 0$, the effective mass in the $\nu$ parameter~\ref{eq:nuterm} becomes $m_{\chi}^2 \rightarrow m_{\chi}^2 + 12 \xi H^2$, and we use the fitting parameters from Table.~\ref{table:beta_k}. As previously discussed, a lighter effective mass during inflation can lead to an IR divergence for modes $k \ll k_{\rm end}$. To explore the effects of small nonminimal coupling, we analyze particle occupation numbers comoving number density spectra for masses $3H_I$ and $4H_I$. We show numerical plots for $n_k$ and $\mathcal{N}_k$ in Figs.~\ref{fig:nkvsk_negxi_1} and \ref{fig:nkvsk_negxi_2} for a range of nonminimal couplings in the range $-1 \leq \xi \leq 0$, evaluated at $a/a_{\rm end} = 100$. These plots illustrate that as $\xi$ becomes more negative, the value of $\nu$ increases, leading to a more red-tilted spectra $\mathcal{N}_k \propto k^{3-2\nu}$. Additionally, the decrease in $\xi$ leads to stronger interference effects, notably for $\xi = -1$ and a mass of $m_{\chi} = 4H_I$.

This phenomenon can be understood analytically as follows. Considering the mode function, $\chi_k$, under the assumption that $\nu$ is imaginary, we can express the mode equation~(\ref{eq:modenu}) for $k \ll aH$ as
\begin{equation}
    \label{eq:modenuim1}
    \chi_k(k \ll a H) \; \simeq \; \frac{1}{2 \sqrt{\pi H}} \left(-\eta H \right)^{3/2} e^{\frac{\pi}{2}\left(\tilde{\nu}-\frac{i}{2}\right)}\left(e^{-\pi \tilde{\nu}} \Gamma(i \tilde{\nu})
    e^{-i \tilde{\nu} \ln(\frac{k}{2aH})}
    +\Gamma(-i \tilde{\nu})e^{i \tilde{\nu} \ln(\frac{k}{2aH})}\right) \,, 
\end{equation}
where $i \nu = \tilde{\nu}$, and the mode function squared becomes
\begin{equation}
    | \chi_k(k \ll a H)|^2 \; = \; - \frac{H^2 \eta^3}{4 \pi} \left(\frac{2\pi \coth{\pi \tilde{\nu}}}{\tilde{\nu}} +  e^{-2i \tilde{\nu} \ln(\frac{k}{2aH})} \Gamma(i \tilde{\nu})^2 +  e^{2i \tilde{\nu} \ln(\frac{k}{2aH})} \Gamma(-i \tilde{\nu})^2  \right) \, .
\end{equation}
The oscillation patterns arise due to interference between the last two terms of opposite phase. These terms can be rewritten as
\begin{equation}
   e^{-2i \tilde{\nu} \ln(\frac{k}{2aH})} \Gamma(i \tilde{\nu})^2 +  e^{2i \tilde{\nu} \ln(\frac{k}{2aH})} \Gamma(-i \tilde{\nu})^2 \; = \; A \cos(2 \tilde{\nu} \log(k/2)) + i B \sin(2 \tilde{\nu} \log(k/2)) \, ,
\end{equation}
where $A = \Gamma(-i \tilde{\nu})^2 + \Gamma(i \tilde{\nu})^2$ is purely real and $B = \Gamma(-i \tilde{\nu})^2 - \Gamma(i \tilde{\nu})^2$ is purely imaginary. Importantly, as $\tilde{\nu}$ increases, the oscillations become more rapid, but their amplitude decreases.
\subsubsection{Large Coupling $\xi < -1 $}
As nonminimal coupling becomes more negative, with $\xi \ll -1$, its effect is two-fold. Firstly, the tachyonic enhancement becomes stonger, leading to more efficient gravitational production during reheating due to broad parametric resonance, which we discussed in Section~\ref{sec:parametricresonance}. Secondly, similar to scenarios with $\xi > 0$, the distribution in the UV region exhibits peaks at higher values of $k/k_{\rm end}$. This shift can be understood from an extended plateau-like region that becomes populated during the reheating phase. As the magnitude of negative nonminimal coupling increases, the resonance strengthens, resulting in greater particle production in the UV region. Consequently, the exponential tail of the spectra shifts to higher $k$ values, occurring when $k^2$ becomes comparable to $a^2 m_{\rm eff}^2$.

\begin{figure}[h!]
\centering
\includegraphics[width=0.8\textwidth]{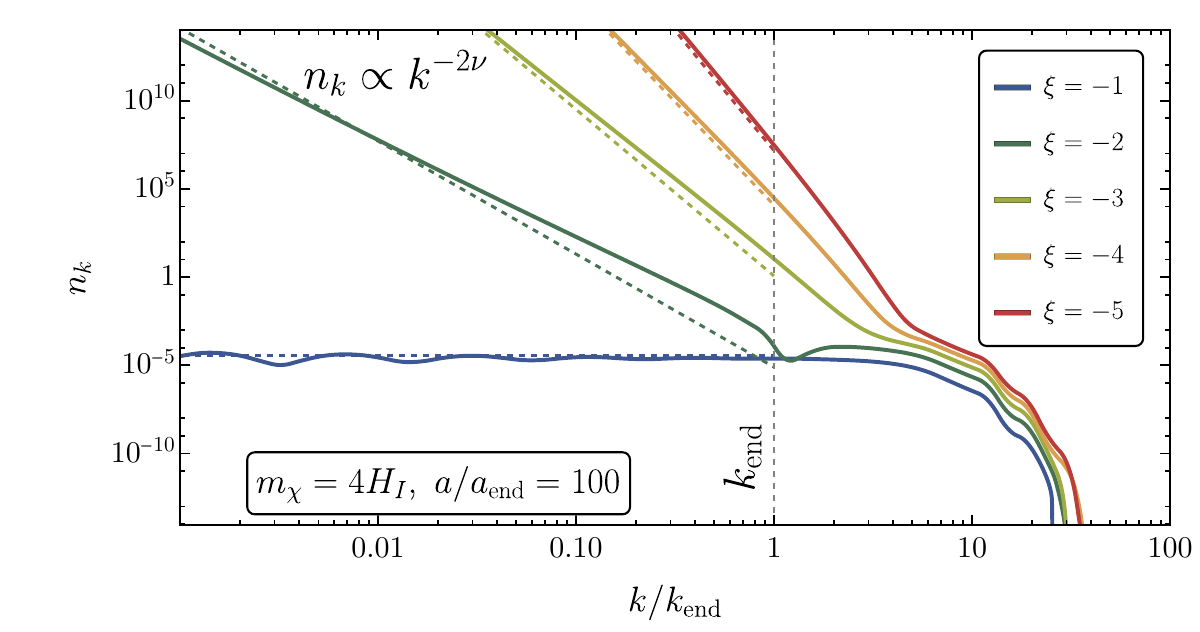}
\includegraphics[width=0.8\textwidth]{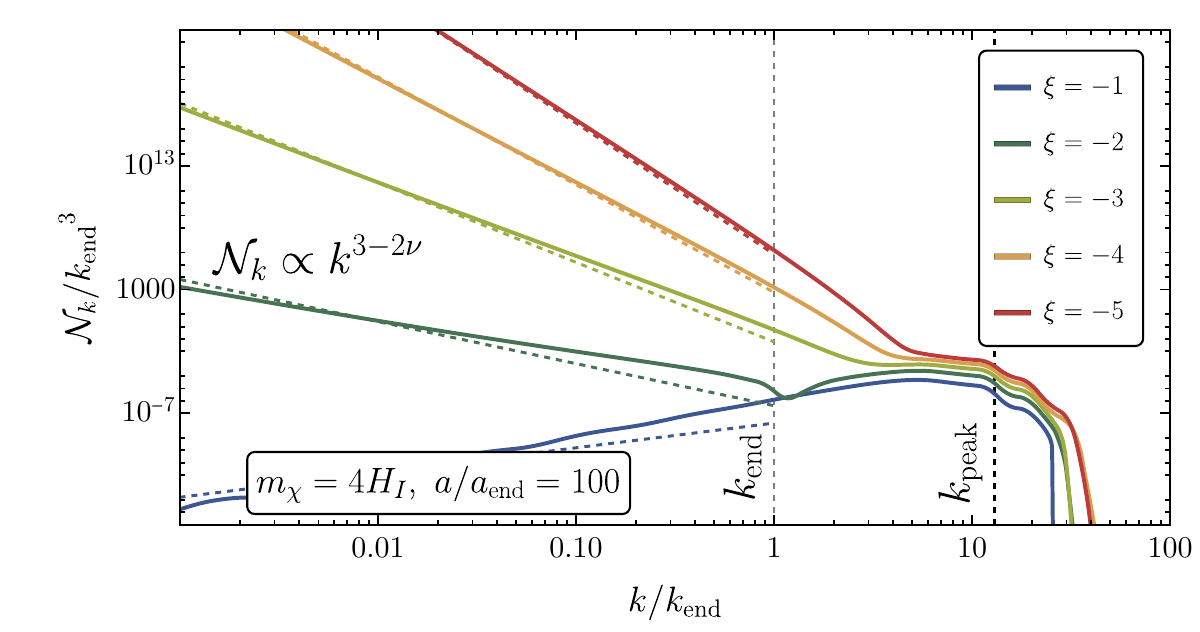}
\caption{Same as Figs.~\ref{fig:nkvsk_negxi_1} and \ref{fig:nkvsk_negxi_2}, but for large negative nonminimal couplings ranging from $-5 \leq \xi \leq -1$.}
\label{fig:xiplotlarge}
\end{figure}

We illustrate the plots for particle occupation number $n_k$ and comoving number density spectrum $\mathcal{N}_k$ for a spectator field mass of $m_{\chi} = 4H_I$ with large negative nonminimal couplings ranging from $-5 \leq \xi \leq -1$ in Fig.~\ref{fig:xiplotlarge}, evaluated at $a/a_{\rm end} = 100$. In the UV region $k > k_{\rm end}$, the spectra peak around $k/k_{\rm end} \sim 13$. As $\xi$ becomes more negative, these peaks shift to even higher values of $k/k_{\rm end}$. Importantly, the IR spectra scales according to the analytical expression derived in Section~\ref{sec:production}, with $n_k \propto k^{-2\nu}$ and $\mathcal{N}_k \propto k^{3-2\nu}$. The UV tail remains exponentially suppressed, yet the UV region displays additional wiggles, which arise due to the oscillating term 
$12\xi R$. This term fluctuates between positive and negative values as the inflaton field oscillates around its minimum. As $\xi$ becomes more negative, not only does the UV region broaden, but it also becomes more densely populated during the post-inflationary gravitational particle production phase.

\subsubsection{Isocurvature constraints}
\label{sec:iso}
In this section, we address the constraints imposed by isocurvature perturbations on superheavy spectator fields with negative nonminimal coupling. 

The current constraints on the isocurvature power spectrum are set by \textit{Planck}, with
\begin{equation}
    \beta_{\rm iso} \; \equiv \; \frac{\mathcal{P}_{\mathcal{S}}(k_*)}{\mathcal{P}_{\mathcal{R}}(k_*) + \mathcal{P}_{\mathcal{S}}(k_*)} < 0.038 \, ,
\end{equation}
at $95 \%$ confidence level, evaluated at the pivot scale $k_* = 0.05 \, \rm{Mpc}^{-1}$~\cite{Planck:2018jri}. Here $\mathcal{P}_{\mathcal{R}}$ denotes the curvature spectra and $\mathcal{P}_{\mathcal{S}}$ represents the isocurvature spectra. During inflation, a spectator field that is light relative to the inflationary scale can lead to a significant production of isocurvature modes, which are in tension with the limits set by \textit{Planck}~\cite{Chung:2004nh}. Assuming $N_* = 55$ $e$-folds of inflation, to avoid excessive isocurvature production, one has to satisfy the limit~\cite{Redi:2022zkt, Garcia:2023awt}
\begin{equation}
    m_{\rm eff} (t_*) \; \geq \; 0.5H_I \; \simeq \; H_{\rm end} \, .
\end{equation}
For minimal coupling $\xi = 0$, this condition simplifies to $m_{\chi} > 0.5H_I$, which is always satisfied for superheavy spectator scalars $m_{\chi} > \frac{3}{2}H_I$. In cases where $\xi > 0$, the effective mass during inflation increases, further suppressing the isocurvature power spectrum.

Applying the effective mass formula $m_{\rm eff}^2  = m_{\chi}^2 + 12 \xi H_I^2$, we can show that the isocurvature constraints are avoided when
\begin{equation}
    \label{eq:isoconstraints}
    \xi > \frac{1}{12} \left( \frac{1}{4} - \frac{m_{\chi}^2}{H_I^2} \right) \, .
\end{equation}
We use this constraint for our dark matter analysis in Section~\ref{sec:dmabundance}.
\subsubsection{Backreaction Effects}
\label{sec:backreact1}
We also briefly discuss the significance of backreaction effects when $\xi < 0$. To account for backreaction, we compare the energy density of the spectator field 	
$\rho_{\chi}$ with that of the inflaton at the end of inflation, with $\rho_{\rm end} = 3H_{\rm end}^2 M_P^2 \simeq 2.9 \times 10^{-11} M_P^4$. Lattice studies suggest that backreaction becomes significant only when the energy density of the spectator field exceeds $1\%$ of the inflaton energy density, with $\rho_{\chi}/\rho_{\phi} > 0.01$. This threshold is incorporated into our analysis of dark matter constraints. However, it is important to note that the isocurvature constraints, discussed in the previous section, generally impose more stringent limits than those from backreaction.
\section{Dark Matter Abundance}
\label{sec:dmabundance}
In this section, we summarize our numerical results and impose dark matter abundance constraints. The present-day dark matter relic abundance is
given by:
\begin{equation}
    \label{eq:dmabundance}
    \Omega_{\chi} h^2\; = \; \frac{\rho_{\chi}(a_0)}{\rho_{c}}h^2\; = \; \frac{m_{\chi} n_{\chi}(a_0)}{\rho_c} h^2 \, ,
\end{equation}
where $\rho_c = 1.054 \times 10^{-5} \, h^2 \, \rm{GeV} \, \rm{cm}^{-3}$ represents the critical energy density today, with $h \sim 0.67$, and $a_0$ is the present-day scale factor. Using the asymptotic value of the comoving number density~(\ref{eq:comovingden}), we can calculate $\Omega_{\chi} h^2$ using the amount of expansion from the end of inflation to today.

The background dynamics are governed by coupled Friedmann-Boltzmann equations:
\begin{align}
\label{eq:dyn1}
\dot{\rho}_{\phi}+3 H \rho_{\phi} =-\Gamma_{\phi} \rho_{\phi} \, , \qquad \dot{\rho}_{r}+4 H \rho_{r} &=\Gamma_{\phi} \rho_{\phi} \, ,  \\
\rho_{\phi}+\rho_{r} =3 M_{P}^{2} H^{2} \, , \qquad
\frac{\diff}{\diff t}{(N w_{\rm int})} &= H w \,, 
\label{eq:dyn4}
\end{align}
where $\rho_r$ is the radiation energy density and $\Gamma_{\phi}$ is the inflaton decay rate. The reheating temperature is defined as the temperature when the inflaton energy density is equal to the energy density of the radiation bath, $\rho_{\phi}(t_{\rm{reh}}) = \rho_r(t_{\rm{reh}})$ at time $t_{\rm reh}$, and is computed from the expression
\begin{equation}
    \label{eq:reheating}
    \rho_r(t_{\rm{reh}}) \; = \; \frac{\pi^2 g_{\rm{reh}} T_{\rm{reh}}^4}{30} \, ,
\end{equation}
where $g_{\rm{reh}}$ represents the effective number of relativistic degrees of freedom at the time of reheating.

Our numerically evaluated values of $n_{\chi}$ contain an infrared divergence for $\xi < 0$, requiring us to account for the duration of reheating to accurately determine the number of $e$-folds of inflation. Assuming entropy conservation post-reheating, the total number of $e$-folds can be expressed as~\cite{Martin:2010kz, Liddle:2003as}:
\begin{equation}
\begin{aligned}
\label{eq:nstarreh}
N_{*} \; = \; \ln \left[\frac{1}{\sqrt{3}}\left(\frac{\pi^{2}}{30}\right)^{1 / 4}\left(\frac{43}{11}\right)^{1 / 3} \frac{T_{0}}{H_{0}}\right]-\ln \left(\frac{k_{*}}{a_{0} H_{0}}\right) -\frac{1}{12} \ln g_{\mathrm{reh}}\\
+\frac{1}{4} \ln \left(\frac{V_{*}^{2}}{M_{P}^{4} \rho_{\mathrm{end}}}\right) +\frac{1-3 w_{\mathrm{int}}}{12\left(1+w_{\mathrm{int}}\right)} \ln \left(\frac{\rho_{\mathrm{rad}}}{\rho_{\text {end }}}\right) 
\, ,
\end{aligned}
\end{equation}
where $H_0 = 67.36\, \rm{km} \, \rm{s}^{-1} \, \text{Mpc}^{-1}$\cite{Planck:2018vyg} is the present Hubble rate and $T_0 = 2.7255 \, \rm{K}$ is the present photon temperature~\cite{Fixsen:2009ug}. 
Here, $\rho_{\rm{end}}$ is the energy density at the end of inflation and $\rho_{\rm{rad}}$ is the energy density at the onset of radiation-dominated era, when $w = p/\rho = 1/3$. In our numerical analysis, we consider the equation of motion parameter over the number of $e$-folds of reheating: 
\begin{equation}
w_{\mathrm{int}} \equiv \frac{1}{N_{\mathrm{rad}}-N_{\mathrm{end}}} \int_{N_{\mathrm{end}}}^{N_{\mathrm{rad}}} w(n) \, \diff n \, ,
\end{equation}
where $N_{\rm{rad}}$ and $N_{\rm{end}}$ indicate the total number of $e$-folds at the end of inflation and the onset of full radiation domination, respectively. 

The IR endpoint of the integral~(\ref{eq:comovingden}) corresponds to the mode that exited the horizon at the start of inflationary epoch. This mode must satisfy the condition $k_{\rm IR}\leq k_0$, where $k_0$ is the present-day horizon scale. This scale can be expressed as~\cite{Martin:2010kz,Liddle:2003as,Ellis:2015pla} 
\begin{equation}
\frac{k_0}{k_{\rm end}} \equiv \frac{a_0 H_0}{a_{\mathrm{end}} H_{\mathrm{end}}}=\left(\frac{90}{\pi^2}\right)^{1 / 4}\left(\frac{11}{43}\right)^{1 / 3}\left(\frac{M_P}{H_{\mathrm{end}}}\right)^{1 / 2} \frac{H_0}{T_0} \frac{g_{\mathrm{reh}}^{1 / 12}}{R_{\mathrm{rad}}},
\end{equation}
with
\begin{equation}
    R_{\rm rad} \; \equiv \; \frac{a_{\rm end}}{a_{\rm rad}} \left(\frac{\rho_{\rm end}}{\rho_{\rm rad}} \right)^{1/4} \; \simeq \; \left(\frac{\Gamma_{\phi}}{H_{\rm end}} \right)^{1/6} \, .
\end{equation}
These expressions are used to compute the IR cutoff. We now proceed to numerically compute the dark matter abundance of gravitationally-produced scalar dark matter, given by Eq.~(\ref{eq:dmabundance}).
\subsection{$\xi > 0$}
\label{sec:dmabundancexipos}
We impose the experimental dark matter constraint $\Omega_{\chi} h^2 = 0.1198$~\cite{Planck:2018vyg}. For scenarios with $\xi > 0$, primary dark matter constraints arise from the reheating temperature constraints, with $T_{\rm BBN} \leq T_{\rm reh} \leq T_{\rm max}$, where the Big Bang Nucleosynthesis (BBN) temperature is given by $T_{\rm BBN} \simeq 4 \, \rm{MeV}$, which is necessary to ensure successful nucleosynthesis~\cite{Kawasaki:2000en, deSalas:2015glj, Hannestad:2004px, Hasegawa:2019jsa},
and $T_{\rm{max}} = \left(90 H_{\rm{end}}^2 M_P^2/\pi^2 g_{\rm{reh}} \right)^{1/4}$, which is the theoretical maximum temperature assuming that all of inflaton energy density is instantaneously converted to radiation at the end of inflation. For the Starobinsky model of inflation, we find $T_{\rm max} \simeq 2.3 \times 10^{15} \, \rm{GeV}$.

We plot the dark matter constraints for $0 \leq \xi \leq 300$ in Fig.~\ref{fig:xiplotlarge}. The blue region is excluded due to constraint $T_{\rm reh} < T_{\rm max}$, and the green region is excluded for $T_{\rm reh} > T_{\rm BBN}$. We note that the $T_{\rm BBN}$ constraint is always more stringent than the backreaction constraint, which we discussed in Section.~\ref{sec:backposxi}. We limit the range up to $\xi = 300$ because it becomes significantly more challenging to numerically obtain clean data for $\xi > 300$ due to increasing numerical noise. For $\xi = 300$, the allowed mass range is $1.6 \times 10^{14} \, {\rm{GeV}} \leq m_{\chi} \leq 3.5 \times 10^{14} \, {\rm {GeV}}$. 

When exploring different values of $\xi$, it is important to remain within the validity of the low-energy theory. We estimate the theory cutoff to be~\cite{Bezrukov:2010jz}
\begin{equation}
    \Lambda \; \simeq \; \frac{M_P}{\rm{max}|\xi|} \, .
\end{equation}
This expression indicates that the temperature of reheating must not exceed this cutoff to ensure the validity of the theory. For the Starobinsky model of inflation, we find $\xi \leq 1000$. After 
carefully cleaning the numerical data, we find that for $\xi = 1000$, the allowed mass range is given by $4.6 \times 10^{14} \, {\rm{GeV}} \leq m_{\chi} \leq 1.2 \times 10^{15} \, {\rm{GeV}}$.

\begin{figure}[h!]
\centering
\includegraphics[width=0.8\textwidth]{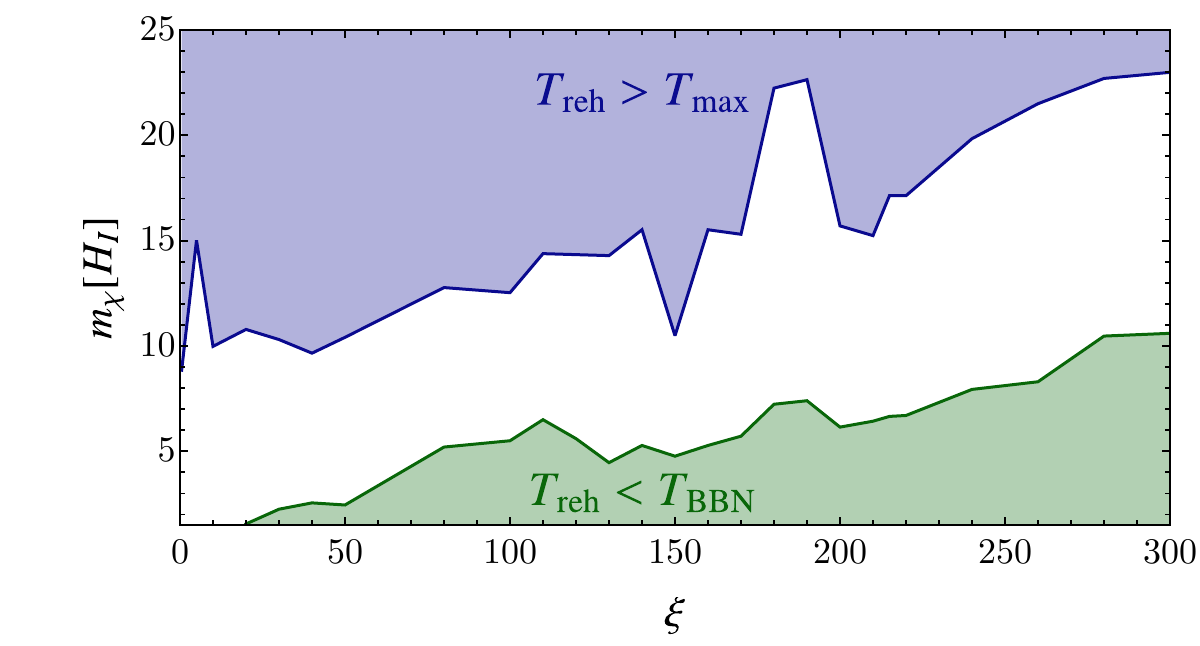}
\caption{Parameter space of the dark matter mass $m_{\chi}$ in units of $H_I = 1.5 \times 10^{13} \, \rm{GeV}$ as a function of the nonminimal coupling $\xi$. The white region indicates the allowed parameter space in agreement with the observed dark matter abundance. The top blue region is excluded by the maximum possible reheating temperature, and the bottom green region is ruled out by the requirement that the reheating temperature must be greater than the BBN temperature.}
\label{fig:xiplotdm}
\end{figure}

\subsection{$\xi < 0$}
\label{sec:dmabundancexineg}
\begin{figure}[h!]
\centering
\includegraphics[width=0.8\textwidth]{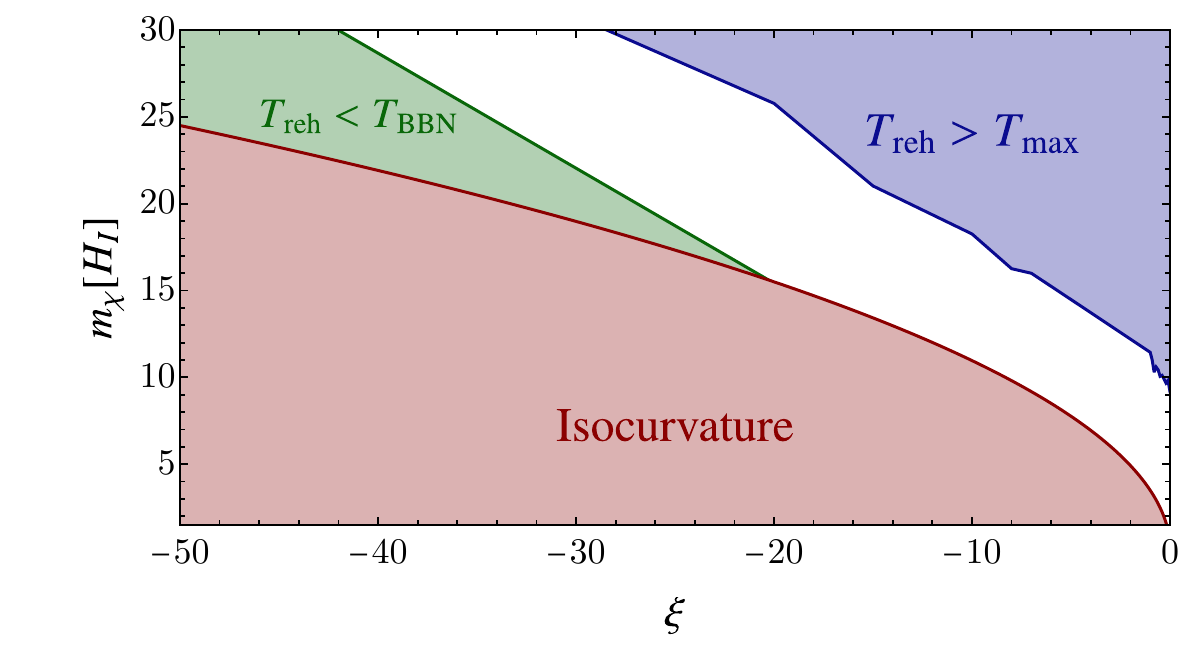}]
\includegraphics[width=0.8\textwidth]{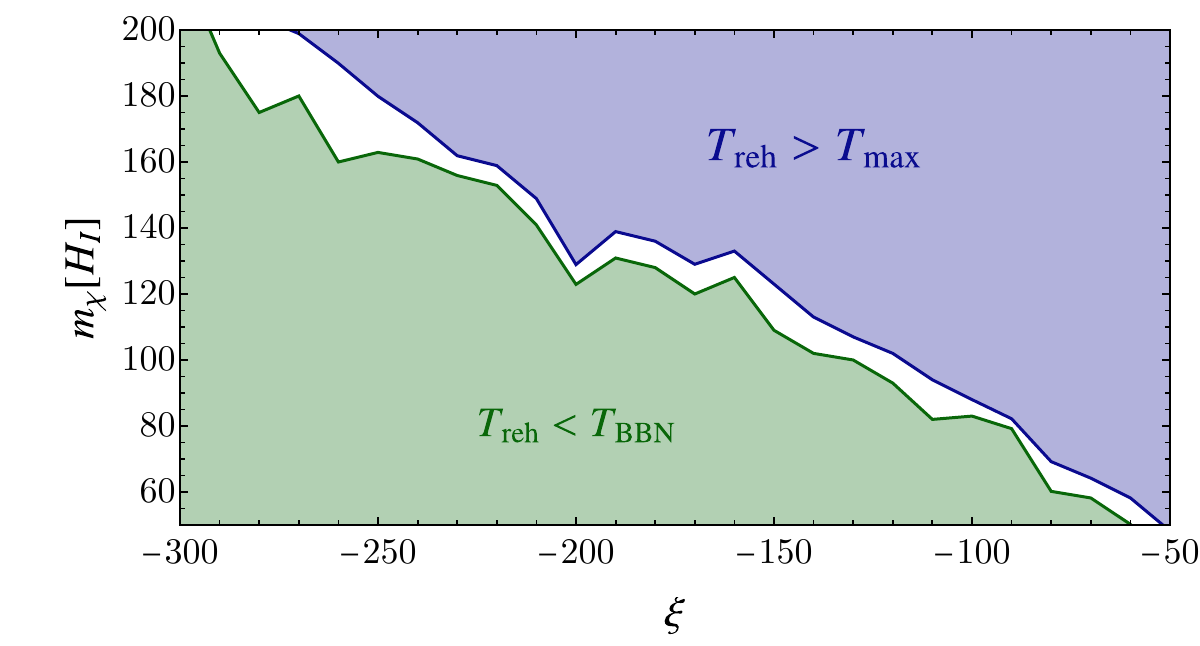}
\caption{Parameter space of dark matter mass $m_{\chi}$ in units of $H_I = 1.5 \times 10^{13} \, \rm{GeV}$ as a function of negative nonminimal coupling $\xi$ in the range of $-50 \leq \xi \leq 0$ (top panel) and $-300 \leq \xi \leq -50$ (bottom panel). The white region shows the allowed parameter space in agreement with the observed dark matter abundance. The top blue region is excluded by the maximum possible reheating temperature, the bottom green region is ruled out because it does not meet the minimum reheating temperature required by BBN constraints, and the red region is excluded by isocurvature constraints.}
\label{fig:xiposdmconst}
\end{figure}

We plot the dark matter constraints for a range of nonminimal coupling values $-300 \leq \xi \leq 0$ in Fig.~\ref{fig:xiposdmconst}. 
Similar to the case with positive $\xi$, we limit the plot to $\xi = -300$ due to substantial numerical noise at more negative values. The blue region is excluded by the constraint $T_{\rm reh} < T_{\rm max}$, the green region is excluded by the constraint $T_{\rm reh} > T_{\rm BBN}$, and the red region is ruled out by the isocurvature constraints, given by Eq.~(\ref{eq:isoconstraints}). 

For $\xi = -300$, the allowed dark matter mass range is $3.3 \times 10^{15} \, {\rm{GeV}} \leq m_{\chi} \leq 3.5 \times 10^{15} \, {\rm{GeV}}$. Although numerical challenges prevent direct evaluation up to $\xi = -1000$, which would align with the theoretical cutoff $\Lambda \simeq T_{\rm max}$, extrapolating the results suggest that reaching $\xi = -1000$ could potentially expand the parameter space up to $\mathcal{O}(10^{16} \, \rm{GeV})$.

\section{Conclusions}
\label{sec:conclusions}
In this paper, we explored the gravitational particle production of superheavy spectator fields with nonminimal coupling, both during and after inflation. We studied different production regimes, including both positive ($\xi > 0$) and negative ($\xi < 0$) nonminimal couplings, and we considered the range $-300 \leq \xi \leq 300$. Our analysis included both analytical and numerical results. For the numerical analysis, we used the Starobinsky model of inflation—a choice strongly supported by current \textit{Planck} constraints on CMB variables.

For $\xi > 0$, As $\xi$ increases, particle production during inflation becomes suppressed, but significantly intensifies after inflation due to efficient parametric resonance. For $\xi < 0$, decreasing $\xi$ results in a red-tilted spectrum in the infrared region. Regardless of $\xi$ sign, the UV tail is always exponentially suppressed for the superheavy mass regime. The qualitative depiction is shown in Fig.~\ref{fig:general}, and the comoving number density spectra is summarized in Fig.~\ref{fig:xi0plots} for $\xi = 0$, in Fig.~\ref{fig:xiposplots2} for $\xi > 0$, and in Fig.~\ref{fig:xiplotlarge} for $\xi < 0$.

The key outcome of this study is the considerable opening of the parameter space allowed by both positive and negative $\xi$, extending up to three orders of magnitude beyond the inflationary scale. For the Starobinsky model of inflation, with $H_I = 1.5 \times 10^{13} \, \rm{GeV}$, the mass range for superheavy dark matter could reach as high as $\sim 10^{16} \, \rm{GeV}$. Our constraints on dark matter are summarized in Fig.~\ref{fig:xiplotdm} for $\xi > 0$ and Fig.~\ref{fig:xiposdmconst} for $\xi < 0$.

This study shows that nonminimal superheavy dark matter (nonminimal WIMPzillas) is a compelling dark matter scenario, allowing such candidates to be as heavy as $\sim 10^{16} \, \rm{GeV}$. While this work focused on the effects and implications of nonminimal coupling, the effects of self-interaction within these models will be explored in future work. The proposed Windchime experiment~\cite{Windchime:2022whs} is sensitive to masses that are close to or larger than the Planck scale, but if improved it could potentially detect masses near the range of heaviest nonminimal superheavy dark matter. We anticipate the upcoming large-scale structure and CMB experiments, which may reveal more about our fundamental understanding of early universe cosmology and nature of dark matter.
\acknowledgments
I would like to thank Marcos A.~G.~Garc{\'i}a, Yohei Ema, Rocky Kolb, Oleg Lebedev, Andrew Long, and Wei Xue. I am also grateful to Dani Figueroa for providing me a version of {\tt ${\mathcal C}$osmo${\mathcal L}$attice} with nonminimal couplings. The work of S.V. was supported in part by DOE grant DE-SC0022148 at the University of Florida.
\appendix
\section{Equations of Motion in Cosmic Time}
\label{app:cosmictime}
In this appendix, we detail the canonical normalization of the mode functions for a scalar spectator field, $\chi$, using cosmic time. The action for the field, $\chi$, is represented as

\begin{equation}
    \label{eq:actionspectscalar}
    \mathcal{S}_{\chi} \; = \; \int d^4 x \sqrt{-g} \left(\frac{1}{2}g^{\mu \nu} \partial_{\mu} \chi \partial_{\nu} \chi 
         - \frac{1}{2}m_{\chi}^2 \chi^2 +\frac{1}{2} \xi \chi^2 R
      \right) \, .
\end{equation}
To canonically normalize the kinetic term, we introduce the following field rescaling
\begin{equation}
   \label{eq:field rescaling}
   X (t, {\bf x}) \; \equiv \; a(t)^{\frac{3}2} \, \chi (t, {\bf x}) \, .
\end{equation}
This rescaling transforms the original action to
\begin{equation}
    \label{eq:spectaction}
    \mathcal{S}_X  \; = \; \int d^4 x \,  {\cal L}_X \; = \; \int d^4 x  \left( 
      \frac{1}{2} \partial^{\mu} X (t, {\bf x}) \partial_{\mu} X  (t, {\bf x})
         -\frac{1}{2} \mu^2(t)\, H^2(t)\,  X^2 (t, {\bf x}) 
      \right) \, ,
\end{equation}
where $H = \dot a / a$ is the Hubble parameter, and we introduced a dimensionless variable,
\begin{equation}
\label{eq:nu_def_cosmictime}
   \mu^2 \; \equiv \; 
   \frac{m_{\chi}^2}{H^2} - \frac{9}{4} + 12 \xi - \left(\frac{3}{2} - 6\xi \right) \frac{\dot{H}}{H^2} \, ,
\end{equation}
where we used $R = -12 H^2 - 6 \dot{H}$. We decompose the field $X$ into its Fourier components:
\begin{equation}
    \label{eq:fourdecomp}
   X (t, \mathbf x) \; = \; 
      \int \frac{d^{3} \mathbf{k}}{ {(2 \pi )^{3}}} \, 
      \left[\hat{a}_{\bf k}^{\,} X_k (t)  \, e^{i {\bf k} \cdot {\bf x} }
       +  \hat{a}^\dagger_{\bf k} X_k^* (t)  \, e^{ - i {\bf k} \cdot {\bf x} } 
      \right] \, .
\end{equation}
Here $\bf{k}$ is the comoving momentum vector, and $\hat{a}^{\dagger}_{\bf k}$ and $\hat{a}_{\bf k}^{\,}$ are the creation and annihilation operators, respectively, satisfying standard commutation relations $[\hat{a}_{\bf k}, \hat{a}_{\bf k'}] = [\hat{a}_{\bf k}^{\dagger}, \hat{a}_{\bf k'}^{\dagger}] = 0$
and $[\hat{a}_{\bf k}^{\,}, \hat{a}_{\bf k'}^{\dagger}] = (2 \pi )^3\delta^{3}(\mathbf{k} - \mathbf{k}')$. 
From Eq.~(\ref{eq:spectaction}), we find that the conjugate momentum of $X$ $\pi \equiv \partial \mathcal{L}_X / \partial \dot{X} = \dot{X}$.
The system is quantized by imposing the commutation relation
\begin{equation}
    [X(t, \mathbf{x}), \pi(t, \mathbf{y})] \; = \; i \delta^{3}(\mathbf{x} -\mathbf{y}) \, ,
\end{equation}
which leads to the Wronskian condition
\begin{equation}
    \label{eq:wronskian1}
    X_k \dot{X}_k^* - X_k^*\dot{X}_k \; = \; i \, .
\end{equation}
The field equation for $X$ in Fourier space can be expressed as
\begin{equation}
    \label{eq:modeeq1}
    \ddot{X}_k(t) + \omega_k^2(t) X_k(t) \; = \; 0 \, ,
\end{equation}
with
\begin{equation}
\label{eq:omega_k}
\omega_k^2 \; = \; \frac{k^2}{a^2} + \mu^2  H^2 \,  .
\end{equation}

To solve the mode equation~(\ref{eq:modeeq1}), we apply the Bunch-Davies vacuum at early times, $t_i \rightarrow -\infty$,
\begin{equation}
    \lim_{t_i \rightarrow - \infty} X_{k} (t_i) \; = \; \frac{1}{\sqrt{2 \omega_k ( t_i)}} e^{-i  \int^{t_i} \omega_k  d t} \, .
   \label{eq:BDvac}
\end{equation}
Finaly, we define the dimensionless occupation number $n_k$ as the energy density of mode $k$ normalized by its frequency~\cite{Kofman:1997yn}:
\begin{equation}
    n_k(t) \;  = \; \frac{1}{\omega_k} \left(\frac 12|\dot{X}_k(t)|^2 +\frac 12\omega_k^2 |X_k(t)|^2 \right) - \frac{1}{2} \, ,
   \label{eq:n_k}
\end{equation}
where we have subtracted the dominant vacuum energy contribution $-\frac{1}{2}$.

\bibliographystyle{JHEP}
\bibliography{refs}

\end{document}